\documentclass[preprint,showpacs,preprintnumbers,amsmath,amssymb, superscriptaddress,longbibliography,nofootinbib]{revtex4-1}
\usepackage{graphicx} 
\usepackage{xcolor}
\usepackage{amsmath}
 \usepackage{hyperref}
\usepackage[top = 2cm, bottom = 2cm, right = 2cm, left =2cm]{geometry}
\usepackage{enumitem}

\usepackage{tikz}
\usetikzlibrary{decorations.pathmorphing,calc}
\begin{document}

\title{Integrable Singularity Black Holes without Inner Horizons in Asymptotically Safe Inspired Gravity and NFW Dark Matter Effects}

\author{Milko Estrada }
\email{milko.estrada@gmail.com}
\affiliation{Departamento de Física, Facultad de Ciencias, Universidad de Tarapacá, Casilla 7-D, Arica, Chile}

\date{\today}

\begin{abstract}
We investigate an alternative ultraviolet (UV) structure for Renormalization Group improved (RG) black holes within a phenomenological framework inspired by asymptotically safe gravity. In several regular RG constructions, regularity at the origin is associated with a dS core and an additional inner horizon, raising issues of spacetime predictability and possible mass inflation instabilities. Motivated by this problem, we relate the RG scale to the radially weighted effective quantum gravitational source entering the volume integrand of the quantum corrected Ricci scalar. This prescription yields a Bessel type running Newton coupling with $G_I(r)\sim r$ in the UV, rather than the usual $G(r)\sim r^3$ behavior, and supports an integrable central singularity. The inner branch is matched through the Israel Darmois conditions to an exterior RG solution with $G_{II}\to G_0$ in the infrared, where $G_0$ is the Newton constant. Within the numerical range investigated, the resulting geometry has a single event horizon and no additional inner horizon. We then incorporate an NFW dark matter distribution coupled to the same running coupling. Since $\rho_{\rm NFW}\sim r^{-1}$ while $G_I(r)\sim r$ near the origin, the effective matter contribution entering the field equations remains finite, producing a cored effective behavior without modifying the NFW profile by hand. The gravitational decoupling algorithm keeps the quantum gravitational and dark matter sectors separately conserved within the same geometry. For the configurations considered, the NFW sector shifts the event horizon while preserving the single horizon structure. Thus, the same running coupling governs both the integrable UV black hole geometry and the softening of the central gravitational response of the NFW source.
\end{abstract}

\maketitle

\section{Introduction}

In recent years, major observational breakthroughs, including the first detection of gravitational waves produced by the coalescence of rotating black holes \cite{LIGOScientific:2016aoc} and the horizon-scale images of the compact objects at the centers of M87 \cite{EventHorizonTelescope:2019dse} and the Milky Way, Sagittarius A* \cite{EventHorizonTelescope:2019ggy}, have provided striking confirmation of several predictions of General Relativity (GR) and compelling evidence for the astrophysical existence of black holes (BHs). At the same time, these achievements have fostered interest in fundamental questions that remain open in the strong-gravity regime and have motivated the development of theoretical proposals beyond the standard classical picture. Along these lines, a particularly challenging direction has been the study of BH physics in the ultraviolet (UV) regime at high-energy scales, where quantum effects could play a relevant role. One of the main motivations for exploring this regime is the presence of singularities at the radial origin of conventional BH solutions in GR. In these regions, curvature invariants typically diverge, signaling the breakdown of the classical description of spacetime. In this context, recent years have seen growing interest in understanding the behavior of BH singularities from a high-energy perspective.

One interpretation of the renormalization group (RG) is to regard it as a mathematical framework describing how the parameters of a theory evolve with the energy scale $k$ at which the system is probed. Although GR is perturbatively non-renormalizable, Weinberg proposed in Ref.~\cite{Weinberg:2009wa} that the gravitational RG flow could approach a non-trivial fixed point in the ultraviolet (UV), potentially allowing for a well-defined quantization of the gravitational field. This idea constitutes the basis of asymptotically safe (AS) gravity. See Appendix~\ref{Apendice} for a brief review. As indicated in Ref.~\cite{Platania:2019kyx}, quantum fluctuations modify the effective dynamics of the theory, making the strength of the gravitational interaction scale dependent. In the infrared (IR) regime, $k\to0$, corresponding to large length scales, one expects to approximately recover the classical description of gravity, whereas the UV  regime, $k\to\infty$, is associated with short distances and increasingly relevant quantum-gravitational effects. A central assumption of AS gravity is that the RG flow approaches a finite and non-trivial ultraviolet fixed point $g_*$. Consequently, the Newton coupling behaves as $G(k)\sim g_*/k^2\to0$ as $k\to\infty$, while its classical value is recovered in the IR, $G(k)\to G_0$ as $k\to0$. The effective gravitational interaction may therefore become weaker at short distances. In black-hole scenarios, this behavior can be combined with an identification between the RG scale and the radial coordinate, $k=k(r)$, modifying the near-origin geometry and potentially rendering curvature invariants finite as $r\to0$, thereby giving rise to regular black-hole (RBH) geometries \cite{Falls:2010he,Platania:2023srt,Bonanno:2000ep}. Along these lines, a commonly considered generalization of the scale setting $k\sim r^{-1}$ is $k\sim1/d_{\rm class}(r)$ \cite{Bonanno:2000ep}, where $d_{\rm class}(r)$ denotes the classical proper distance and behaves as $d_{\rm class}(r)\sim r^{3/2}$ in the UV and $d_{\rm class}(r)\sim r$ in the IR. This prescription yields $G(r)\big|_{r\to0}\sim r^3$, while $G(r)\big|_{r\to\infty}\sim G_0$. Such a UV behavior can soften or remove the classical black-hole singularity. Asymptotically safe gravity has attracted increasing attention in recent years as a framework for describing regular black holes, as discussed in Refs. \cite{Hassannejad:2024cbu,Pawlowski:2023dda,Kanai:2026sdu,Shi:2023kid,Stashko:2024wuq,Bonanno:2023rzk}. 

From a phenomenological point of view, the RG energy scale can be related to curvature invariants, since the spacetime geometry itself provides characteristic physical scales for quantum fluctuations of the gravitational field \cite{Pawlowski:2018swz,Borissova:2022jqj}. Regions of stronger curvature are naturally associated with shorter length scales and, consequently, with higher energies, where quantum-gravitational effects are expected to become more relevant. This provides a geometric basis for establishing a correspondence between the gravitational field and the RG scale $k$. Along these lines, Ref.~\cite{Ishibashi:2021kmf} employs the Kretschmann invariant to define the energy scale from the classical geometry. This choice is particularly appropriate for vacuum solutions such as Schwarzschild, where the Ricci tensor vanishes while the Kretschmann invariant remains nonzero and still captures the strength of the tidal field. A related identification based on the Kretschmann scalar was proposed in Ref.~\cite{Pawlowski:2018swz} to investigate quantum modifications of black-hole geometries. Ref.~\cite{Borissova:2022jqj}, in turn, emphasizes that selecting a local geometric scale constructed from curvature invariants amounts to a physical hypothesis about the quantities governing the onset of quantum effects. Alternative prescriptions involving the Ricci scalar have also been explored. Ref.~\cite{Bonanno:2020bil} points out that identifying the RG scale with the Ricci scalar, together with the ultraviolet behavior of the gravitational couplings, leads to an effective action proportional to $R^2$, consistently with the expected scaling near the fixed point. In this context, see also Ref.~\cite{Platania:2019qvo}. Taken together, these works support the use of curvature-dependent RG scale identifications as phenomenological prescriptions for incorporating possible quantum effects through geometry-dependent gravitational couplings. However, the precise relation between $k$ and the curvature invariants is not uniquely determined from first principles of asymptotically safe gravity; rather, it depends on the relevant physical scales, the spacetime geometry under consideration, and the particular procedure adopted to implement the quantum corrections \cite{Bonanno:2020bil,Borissova:2022jqj}. An interesting application is developed in Ref.~\cite{Platania:2019kyx}, where the RG energy scale is related to quantum corrections to curvature invariants through an iterative procedure in which the running Newton coupling converges to a form associated with the Dymnikova solution \cite{Dymnikova:1992ux}. These curvature invariants acquire quantum corrections characterized by a scale of order $\sim G_0,\rho_{\rm qc}(r)$, where $\rho_{\rm qc}$ represents the effective quantum-gravitational energy density associated with the running gravitational coupling, which may counteract the classical tidal contribution and contribute to regulating curvature divergences.

On the other hand, several regular black-hole (RBH) solutions obtained through RG-improved running gravitational couplings in asymptotically safe gravity lead to geometries characterized by an additional inner horizon associated with a de Sitter (dS) core, see for example \cite{Falls:2010he,Bonanno:2000ep,Platania:2019kyx}. The presence of such an inner horizon may introduce further difficulties in the resulting causal structure. In particular, the inner horizon is susceptible to the well-known mass-inflation mechanism, which can produce strong dynamical instabilities \cite{Poisson:1990eh,Brown:2011tv}. Subsequent analyses have emphasized that this instability represents a serious issue for non-singular black-hole geometries possessing an inner horizon with non-vanishing surface gravity \cite{Carballo-Rubio:2022pzu}. Moreover, since such an inner horizon can also play the role of a Cauchy horizon, its presence raises questions concerning the predictability of spacetime beyond this surface \cite{Ovalle:2023vvu}. Against this background, an alternative possibility that has attracted attention consists of replacing the regular core by an integrable singularity (IS) \cite{Lukash:2013ts}. In this class of configurations, curvature quantities may diverge at the origin while the corresponding volume integral of the Ricci scalar remains finite. In particular models, such singularities can be weak according to Tipler's classification \cite{Nolan:2000rn}, implying that the integrated tidal distortion need not become destructive and that radial geodesics may admit an extension through the singular region \cite{Estrada:2023dcj}. Possible connections between integrable singularities and quantum-gravitational effects have also been investigated \cite{Casadio:2023iqt}. Further applications and physical developments of this framework can be found in Refs.~\cite{Ovalle:2023vvu,Estrada:2024moz,Arrechea:2025fkk,Estrada:2026kea}.

Motivated by these considerations, the first problem we address concerns whether an RG-improved running gravitational coupling, within a framework motivated by AS gravity, can support an alternative UV black-hole structure characterized by an integrable singularity while avoiding the additional inner horizon associated with a potentially unstable dS core. From this perspective, the identification of the RG scale plays a central role. Rather than adopting a scale-setting prescription primarily aimed at rendering curvature invariants finite, we consider a phenomenological identification motivated by the radial integrability of the curvature near the origin, thereby allowing the central divergence to persist in an integrable form. In this framework, the effective quantum-gravitational source is characterized through its radially weighted contribution to the curvature, providing an alternative criterion for describing the short-distance regime. This distinction allows an UV behavior of the running Newton coupling different from the cubic scaling commonly associated with regular dS cores in RG-improved black-hole geometries. The resulting UV branch is required to remain finite and compatible with a real RG scale within its domain of validity. At short distances, the corresponding running Newton coupling acquires a non-trivial Bessel-type radial dependence, leading to a linear behavior near the origin rather than the usual cubic scaling. Since this short-distance branch is not intended to describe the entire spacetime, it is smoothly connected to an outer running-coupling solution that approaches the classical Newton constant in the IR. The matching between both regions is constrained by the corresponding junction conditions, allowing the UV construction to be consistently embedded into a black-hole geometry extending toward large distances. For the configurations considered below, the resulting spacetime exhibits a single event horizon, while preserving the expected recovery of the classical gravitational coupling in the IR.

A second problem we address arises when an additional matter source, specifically a dark matter (DM) distribution, is incorporated into the system. The NFW profile \cite{Navarro:1995iw} provides a widely used description of DM halos, but it develops a central cusp with increasing density toward the origin, diverging as $\sim r^{-1}$, in contrast with the approximately cored behavior inferred for several galaxies, giving rise to the well-known cusp--core problem \cite{Deng:2018jjz}. From a quantum perspective, Ref.~\cite{Deng:2018jjz} investigates whether ultralight bosonic DM can generate a central core through quantum effects, although the resulting core properties do not generally reproduce observations. A different quantum approach has been explored in Ref.~\cite{Musielak:2024yww}, where an NFW halo is embedded in a model leading to a central core surrounded by an NFW envelope with quantized orbits. In recent years, BH solutions of the Einstein equations surrounded by NFW galactic halos have attracted increasing attention; see, e.g., Ref.~\cite{Xu:2018wow}, the references therein, and subsequent applications. In this context, it is natural to ask how the gravitational response of an NFW distribution is modified when incorporated into our RG-improved scenario. In a conventional description, the DM source couples through the constant Newton coupling $G_0$, whereas in our construction the running coupling $G(r)$ introduces a radial weighting of the source. The gravitational equations can therefore be interpreted as defining an effective density whose radial structure emerges from the coupling to $G(r)$ rather than from an ad hoc modification of the original NFW profile. Since, as can be seen below, the running coupling exhibits a linear UV behavior near the origin, $G(r)\big|_{r\to 0}\sim r$, we investigate whether this mechanism can convert the effective central cusp into a finite behavior. In this way, a direct link is established between the UV structure of the BH geometry and the central behavior of the DM distribution, without assuming that this mechanism by itself provides a complete solution to the cusp--core problem at galactic scales, which lies beyond the scope of the present work. To incorporate this additional source without modifying the RG prescription established above, we employ the gravitational decoupling (GD) \cite{Ovalle:2017fgl} method through its minimal geometric deformation (MGD) implementation. This procedure separates the quantum-gravitational sector generating the seed geometry from the additional contribution associated with the NFW profile, while keeping both sectors separately conserved, consistently with the assumption that the DM sector does not directly exchange energy and momentum with the seed source and interacts with it only through the gravitational field. It also allows us to examine how the new source deforms the geometry and modifies its horizon structure relative to the configuration without DM. In particular, we investigate whether, after including this additional contribution, the spacetime preserves a single-horizon structure and remains free of a potentially unstable dS core and its associated inner horizon.

\section{Our Problem: Searching for an Integrable Singularity in Asymptotically Safe Gravity}

In Ref.~\cite{Platania:2019kyx}, assuming an equation of state of the form $p_{\rm qc}=\omega,\rho_{\rm qc}$, the quantum-corrected Ricci scalar is written as
\begin{equation}
R_{\rm qc}(r)=
\frac{G_{\rm qc}}{G_0}
\left[
\bar{R}_{\rm cl}
+
16\pi(1-\omega)\,
G_0\,\rho_{\rm qc}(r)
\right],
\label{eq:Ricci_n}
\end{equation}
where $\bar R_{\rm cl}=0$ for classical Schwarzschild spacetime, while $\bar K_{\rm cl}$ diverges at the origin, indicating divergent tidal forces. The scale $\sim G_0\,\rho_{\rm qc}(r)$ characterizes quantum curvature corrections that may oppose classical tidal forces and regulate curvature divergences. This motivates $k^2(r) \equiv\mathcal{K}[R_{\rm qc}]\sim G_0\,\rho_{\rm qc}(r)$, yielding the converged running coupling equation
\begin{equation}
\label{recursiveG}
G_{\rm qc} \equiv G(r)=
\frac{G_0}
{
1+g_*^{-1}G_0\,
\mathcal{K}[R_{\rm qc}(r)]
}.
\end{equation}
where the effective energy density can then be interpreted as a quantum-gravitational self-energy responsible for the possible weakening of tidal forces and regularization of the spacetime.

In this work, rather than seeking a convergent iterative procedure, we aim to identify a suitable phenomenological prescription for the RG energy scale to investigate whether an integrable singularity can be characterized near the origin in the high-energy regime, while suppressing the formation of a potentially unstable inner horizon associated with a de Sitter core. We begin by rewriting the expression for the quantum-corrected Ricci scalar in an asymptotically safe background, Eq.~\eqref{eq:Ricci_n}, as

\begin{equation}
\label{eq:Ricci_n1}
R_{\rm qc}
=
G_{\rm qc}
\left[
\frac{\bar{R}_{\rm cl}}{G_0}
+
(\mbox{EQC})
\right].
\end{equation}
where, in natural units, $[R_{\rm qc}]=[\bar{R}_{\rm cl}]=\ell^{-2}$ and $[G_{\rm qc}]=[G_0]=\ell^{2}$, as for the usual Newton constant, with $\ell$ denoting a length scale. Thus, $[(\mbox{EQC})]=\ell^{-4}$. Here, using the equation of state $p_{\rm qc}=\omega \,\rho_{\rm qc}$, the effective quantum contribution (EQC) is defined as

\begin{equation}
\label{EQC}
(\mbox{EQC})
=
16\pi(1-\omega)\,
\rho_{\rm qc}(r).
\end{equation}
where $\omega \equiv \omega(r)$. We assume that this effective quantum contribution is sourced by an effective energy-momentum tensor of the form

\begin{equation}
\label{TEQC}
\left(T^\mu{}_\nu\right)_{\rm qc}
=
\operatorname{diag}
\left(
-\rho_{\rm qc},
-\rho_{\rm qc},
p_{\rm qc},
p_{\rm qc}
\right),
\end{equation}
where $[\rho_{\rm qc}]=[p_{\rm qc}]=\ell^{-4}$, while $\omega$ is dimensionless. The quantities $\rho_{\rm qc}$ and $p_{\rm qc}$ should be understood as effective quantum-gravitational density and pressure, rather than ordinary matter variables. They encode the effective source associated with the running of the Newton coupling and the corresponding backreaction on the geometry. The trace is therefore

\begin{equation}
\label{TraceEQC}
T_{\rm qc}
=
-2(1-\omega)\,
\rho_{\rm qc}
=
-2\,\rho_{\rm qc}
+
2\,p_{\rm qc}.
\end{equation}

It then follows directly from Eqs.~\eqref{EQC} and \eqref{TraceEQC} that

\begin{equation}
\label{EQCTrace}
(\mbox{EQC})
=
-8\pi\,
T_{\rm qc}.
\end{equation}

Thus, in a consistent way, Eq.~\eqref{eq:Ricci_n1} can be written as

\begin{equation}
\label{eq:Ricci_n2}
R_{\rm qc}
=
G_{\rm qc}
\left[
\frac{\bar{R}_{\rm cl}}{G_0}
-
8\pi\,
T_{\rm qc}
\right].
\end{equation}

The corresponding static and spherically symmetric line element is written as
\begin{equation}
\label{MetricEQC}
ds^2
=
-f(r)\,dt^2
+
\frac{dr^2}{f(r)}
+
r^2d\Omega^2,
\end{equation}
with
\begin{equation}
\label{fEQC}
f(r)
=
1-\frac{2mG(r)}{r}.
\end{equation}

In agree with the above discussion, we seek a radial dependence $G_{\rm qc}\equiv G(r)$ to investigate the possibility of an integrable singularity in the high-energy regime.For this geometry, the relevant mixed components of the Einstein tensor are
\begin{equation}
\label{GttEQC}
\mathcal{G}^{t}{}_{t}
=
-\frac{2mG'(r)}{r^2},\,\,\,\mathcal{G}^{\theta}{}_{\theta}
=
-\frac{mG''(r)}{r}.
\end{equation}
where $[m]=\ell^{-1}$ is a constant related to the usual geometrized mass term through $[m_{\rm geo}]=[m]\,[G(r)]=\ell$. We can directly verify that, consistently with the dimensions of the quantum-corrected and classical Ricci scalars, $[\mathcal{G}^{\mu}{}_{\nu}]=\ell^{-2}$. Following \cite{Platania:2019kyx}, the corresponding effective Einstein equations are
\begin{equation}
\label{EinsteinEQC}
\mathcal{G}^{\mu}{}_{\nu}
=
8\pi \, G(r)\,
\left(T^\mu{}_\nu\right)_{\rm qc}
\end{equation}
Thus, we obtain the effective quantum-gravitational density and pressure components
\begin{equation}
\label{rhoEQC}
\rho_{\rm qc}(r)
=
\frac{mG'(r)}
{4\pi r^2G(r)}\,\,\,, \,\,\, 
p_{\rm qc}(r)
=
-\frac{mG''(r)}
{8\pi rG(r)}.
\end{equation}

Thus, the trace of the effective energy-momentum tensor is given by
\begin{equation}
\label{TraceEQCr}
T_{\rm qc}(r)
=
-\frac{m}{4\pi } \,
\left[\frac{2G'(r)+rG''(r)
}{r^2 G(r)}.\right]
\end{equation}

\subsection{Integrable Singularity at Short-Length Scales}

In the literature, integrable singularities have been considered for which the Ricci scalar may diverge near the origin, while its contribution integrated over an infinitesimal volume remains well defined, namely, $\int_0^\epsilon r^2 R(r)\,dr < \infty$.
We shall focus on the simplest case in which, even though $R(r)$ may be singular at $r=0$, the geometry exhibits an integrable singularity because the radially weighted curvature $r^2R(r)$ remains sufficiently regular near the origin, i.e. in the short-distance/high-energy regime, for the above integral to converge. Along these lines, we point out that Eq.~\eqref{eq:Ricci_n2} can be written as follows:
\begin{equation}
\label{r2Ricci}
\underbrace{r^2R(r)}_{\text{radially integrated curvature}}
=
G(r)
\left[
\underbrace{\frac{r^2\bar R_{\rm cl}}{G_0}}_{\text{classical sector}}
-
\underbrace{8\pi r^2\,T_{\rm qc}(r)}_{\text{quantum-gravitational sector}}
\right].
\end{equation}

We assume that the effective Newton coupling $G(r)$ is a regular function, free of singularities. This assumption is consistent with the physical scenario explored here, in which the running of the gravitational coupling is expected to soften, or eventually remove, the pathological behavior of the classical geometry at high-energy scales, associated with the presence of a potentially unstable inner horizon arising from a dS core. Consequently, it is natural to require that the contribution associated with the effective quantum-gravitational source be radially well behaved by itself, namely, that $\sim G(r) \, r^2\,T_{\rm qc}(r)$ does not generate a new non-integrable divergence at the origin. 

Since our analysis is based on the radial integrability of the curvature near the origin, we focus on the structure of the corresponding radial curvature integrand. As can be seen from Eq.~\eqref{r2Ricci}, the factor $G(r)$ multiplies both the classical and the effective quantum-gravitational sectors. In natural units, $r^2R(r)$ is dimensionless, and the effective quantum-gravitational contribution to this quantity is given by the product $G(r)r^2T_{\rm qc}(r)$. We therefore separate the common gravitational coupling $G(r)$ and consider the radially weighted effective source $r^2T_{\rm qc}(r)$ as the quantity characterizing the strength of the quantum-gravitational sector.
\begin{equation} \label{IdentificacionMia}
k^2(r)=\mathcal{K}
=
-\xi\,4\pi r^2T_{\rm qc}(r).
\end{equation}
where the factor $4\pi$ originates from the angular integration in spherical symmetry and $\xi$ is a positive dimensionless parameter. We further point out that this identification is dimensionally consistent. Since $[k^2]=\ell^{-2}$, while $[\xi]=\ell^0$, $[r^2]=\ell^2$, and $[T_{\rm qc}(r)]=\ell^{-4}$, it follows that $[\xi \, r^2\,T_{\rm qc}(r)]=\ell^{-2}$, as required.  We therefore use $-4\pi r^2T_{\rm qc}(r)$ to characterize the strength of the effective quantum-gravitational source relevant for the scale identification in the short-distance regime. As discussed below, in this work we restrict ourselves to the high-energy regime in which $k^2(r)>0$, thereby avoiding imaginary values of the RG scale $k$. This prescription is conceptually different from the scale identification introduced in Ref.~\cite{Platania:2019kyx}. In that construction, the identification $k^2=\xi G_0\rho$, naturally emerging from the quantum-improved curvature invariants, characterizes the strength of quantum-gravitational effects. In the present case, instead, the scale identification is motivated by the radial integrability of the curvature near the origin. Thus, rather than being determined by a purely local density scale, the RG scale is associated with the radially weighted effective source $4\pi r^2T_{\rm qc}(r)$, which characterizes the quantum-gravitational contribution entering the corresponding integrability condition.

\subsection{Our Set--up}

First, from Eqs. \eqref{GUV} and \eqref{recursiveG}, we note that
\begin{equation}
k^2(r)=\frac{g_*}{G_0}\left(\frac{G_0}{G(r)}-1\right).
\end{equation}

Therefore, to avoid imaginary values of $k$, we restrict our search for an integrable singularity to a radial domain where $G(r)<G_0$, corresponding to short length scales and high energies. Accordingly, we consider a running coupling $G(r)$ such that the trace in Eq.~\eqref{IdentificacionMia} satisfies $T_{\rm qc}(r)<0$ within this domain. Once the form of $G(r)$ has been determined, we will discuss the corresponding restrictions on the radial domain.

As will be seen below, the imposed restrictions lead to a Bessel-type running coupling $G(r)$, which is monotonically increasing, finite, and satisfies $0<G(r)<G_0$ within a radial domain $0<r<R_c$. The value of $R_c$ relevant to the scenario considered here will be analyzed numerically below. Beyond this domain, the solution may exceed $G_0$ and exhibit oscillatory behavior, making its extension incompatible with the scale identification adopted here. We therefore consider the following piecewise construction:

\begin{itemize}
\item We define the inner radial domain $0\leq r<R_I$ as {\it Region I}. As we will verify below, $R_I<R_c$ is required in order to satisfy the Israel junction conditions at $r=R_I$. Within this region, we seek a running coupling $G(r)=G_I(r)$ that characterizes an integrable singularity near the radial origin in the short-distance/high-energy regime. As we will verify below, our scale identification in Eq. \eqref{IdentificacionMia} is valid within this domain, whereas its extension to larger radii may lead to negative values of $k^2$. In this regard, as we will test below, the corresponding trace in Eq. \eqref{TraceEQCr} changes sign beyond the selected domain, making the scale identification incompatible with real values of $k$. This behavior can be physically understood in terms of the decreasing effective energy density with increasing radius, accompanied by a growing relative contribution of the pressures to the trace. 

\item The outer radial domain, $R_I\leq r<\infty$, is referred to as {\it Region II}. In this region, we introduce a different functional form of $G_{II}(r)$ that approaches the classical Newton constant, $G_{II}(r)\to G_0$, as $r\to\infty$. 

\item We therefore consider the following piecewise construction,
\begin{equation} \label{tramos}
G(r)=
\begin{cases}
G_I(r), & 0\leq r\leq R_I,\\[1mm]
G_{II}(r), & r\geq R_I,
\end{cases}
\end{equation}
where the two branches are required to be consistently matched at $r=R_I$
As we will test below, the corresponding geometries are required to satisfy the Israel junction conditions at $r=R_I$.
\end{itemize}

\subsection{Our Improved Effective Running Gravitational Coupling in Region I}
In agreement with the discussion above, our scale identification is intended to describe the short-distance regime. We therefore begin by determining the corresponding running coupling in Region I and subsequently analyze the numerical values of the upper bound $r=R_I$ compatible with the matching conditions. In this way, considering $G(r)=G_I(r)$ and taking into account Eqs.~\eqref{IdentificacionMia} and \eqref{TraceEQCr}, the relation \eqref{recursiveG} leads to
\begin{equation}
\label{Glimite1}
G_I(r)
=
\frac{G_0}
{
1+
\displaystyle 
A\,\frac{2G_I'(r)+r\,G_I''(r)}{G_I(r)}
}.
\end{equation}
where
\begin{equation} \label{EcuacionA}
    A=\frac{\xi m G_0}{g_*}
\end{equation}

Equation \eqref{Glimite1} can be rewritten as
\begin{equation}
rG_I''(r)+2G_I'(r)
+\frac{1}{A}\left[G_I(r)-G_0\right]
=0.
\end{equation}
Defining
\begin{equation} \label{RelacionG-F}
 G_I(r)=G_0+ F(r) \Rightarrow F(r)=G_I(r)-G_0, \end{equation}
we obtain
\begin{equation}
rF''(r)+2F'(r)+\frac{1}{A}F(r)=0.
\end{equation}
To solve this differential equation, we proceed as follows. First, we introduce the auxiliary variable
\begin{equation} \label{CoordenadaZR}
z
=
2\sqrt{\frac{r}{A}}.
\end{equation}

In terms of this new variable, the differential equation becomes
\begin{equation} \label{EqDiferencialF}
F_{zz}
+
\frac{3}{z}F_z
+
F
=
0,
\end{equation}
where $F_z\equiv dF/dz$ and $F_{zz}\equiv d^2F/dz^2$. Next, we perform the redefinition
\begin{equation} \label{SolucionBesselF}
F(z)
=
\frac{Y(z)}{z}.
\end{equation}

Substituting this expression into the previous equation \eqref{EqDiferencialF}, we obtain
\begin{equation}
z^2Y_{zz}
+
zY_z
+
\left(z^2-1\right)Y
=
0,
\end{equation}
where $Y_z\equiv dY/dz$ and $Y_{zz}\equiv d^2Y/dz^2$. This is the ordinary Bessel equation of order one, whose general solution is
\begin{equation} \label{SolucionBesselY}
Y(z)
=
C_1J_1(z)
+
C_2Y_1(z),
\end{equation}
where $J_1(z)$ is the Bessel function of the first kind of order one, $Y_1(z)$ is the Bessel function of the second kind of order one, and $C_1$ and $C_2$ are integration constants. Substituting Eqs.~\eqref{SolucionBesselY}, \eqref{SolucionBesselF}, and \eqref{CoordenadaZR} into Eq.~\eqref{RelacionG-F}, we obtain

\begin{equation}
G_I(r) \label{ConstanteG(r)1}
=
G_0+
C_1\,\frac{
J_1\!\left(2\sqrt{r/A}\right)}{2\sqrt{r/A}}
+
C_2\,\frac{Y_1\!\left(2\sqrt{r/A}\right)
}{
2\sqrt{r/A}
}.
\end{equation}

We first note that the contribution proportional to $Y_1(z)$ in the second term of the last equation diverges as $r\rightarrow0$. Indeed, for $Y_1(z)\big|_{z\rightarrow 0}\sim -\frac{2}{\pi z}$, and therefore $\frac{Y_1(z)}{z}\big|_{z\rightarrow 0}
\sim
-\frac{2}{\pi z^2}
\sim
\frac{1}{r}$. Since, in agreement with the discussion above, we seek a non-divergent expression for $G_I(r)$ at the origin, we must impose $C_2=0$, so that the second term in the last equation is suppressed. 

On the other hand, in agreement with the discussion above, we require the running Newton coupling to become small in the high-energy UV regime, corresponding to short distances. Thus, we impose $G_I(0)\sim 0$. We use the fact that, for $r\to0$, equivalently $z\to0$, one has $J_1(z)/z\to1/2$. Evaluating this limit in Eq.~\eqref{ConstanteG(r)1}, we obtain $C_1=-2G_0$. Therefore, the running Newton coupling, Eq.~\eqref{ConstanteG(r)1}, is given by
\begin{equation} \label{G1Expresion}
G_I(r)
=
G_0
\left[
1-
\frac{
J_1\!\left(2\sqrt{r/A}\right)
}{
\sqrt{r/A}
}
\right].
\end{equation}

\subsection{Behavior of $G_I(r)$ at short length scales}

At short length scales, we point out the following:
\begin{itemize}
    \item First, to study the behavior near the origin, we use $J_1(z)=z/2-z^3/16+z^5/384+\cdots$. With $z=2\sqrt{r/A}$, we obtain $J_1\left(2\sqrt{r/A}\right)/\sqrt{r/A}=1-r/(2A)+r^2/(12A^2)+\cdots$. Therefore, $G_I(r)=G_0\left[r/(2A)-r^2/(12A^2)+\cdots\right]$. Thus, in the short-distance regime, the term $\sim r$ becomes dominant, yielding
\begin{equation} \label{GIOrigen}
    G_I(r)\,\big|_{r \to 0} \sim \frac{G_0}{2A} \, r
\end{equation}

Hence, $G_I(r)\sim r$ in the UV. This already marks a qualitative difference from the usual behavior $G(r) \big|_{r \to 0}\sim r^3$ reported in the literature, for example in Refs.\cite{Bonanno:2000ep, Platania:2019kyx,Platania:2023srt}, which commonly leads to the formation of a de Sitter core and may be accompanied by an inner horizon potentially subject to instability. Below, we will investigate whether the present UV behavior is able to suppress the formation of such an inner horizon associated with a potentially unstable de Sitter core

\item It is worth briefly discussing the Planckian scenario. We can introduce the characteristic length scale $\ell_{\rm cr}\equiv A=(\xi/g_*)\,m_{\rm geo}$, where $m_{\rm geo}\equiv m\,G_0$ is the usual geometrized mass, with $[m_{\rm geo}]=\ell$. Therefore, the running Newton coupling depends on the dimensionless ratio $r/\ell_{\rm cr}$. For a potential Planckian configuration, $m\sim\ell_{\rm p}^{-1}$ and $G_0\sim\ell_{\rm p}^{2}$ imply $m_{\rm geo}\sim\ell_{\rm p}$, where $\ell_{\rm p}$ denotes the Planck length. Hence, $\ell_{\rm cr}\sim(\xi/g_*)\,\ell_{\rm p}$. If the finite ratio $\xi/g_*$ is sufficiently larger than unity, one may have a hypothetical physical situation in which $r\sim\ell_{\rm p}\ll\ell_{\rm cr}$. For a Planckian configuration, the system then lies in the UV regime, where the running of the gravitational coupling becomes significant and quantum-gravitational modifications may become relevant at distances of order $\ell_{\rm p}$ and below.
\end{itemize}

\subsection{Radial Domain of Region I}

Given the considerations above, we point out the following

\begin{itemize}
    \item $G_I(0)=0$, and, in agreement with Eq.~\eqref{GIOrigen}, the running coupling grows linearly in the vicinity of the origin.
    \item In agreement with the discussion above, we require $0\leq G_I(r)<G_0$ in Region I in order to avoid imaginary values of the RG scale $k$. To determine the radial domain in which this condition is satisfied, we introduce 
    \begin{equation} \label{RelacionXR}
        x=2\sqrt{\frac{r}{A}}\geq0
    \end{equation}
    Then, from equation \eqref{G1Expresion}, the running coupling can be written as
\begin{equation} \label{Ratio1}
\frac{G_I(r)}{G_0}
=
1-\frac{2J_1(x)}{x}.
\end{equation}

The upper condition $G_I(r)<G_0$ requires $G_I(r)/G_0<1$, which implies $1-2J_1(x)/x<1$. Therefore, for $x>0$, we must have $J_1(x)>0$. The Bessel function of the first kind  of order one $J_1(x)$ is positive between the origin and its first positive zero, denoted by $j_{1,1}$, where $J_1(j_{1,1})=0$ and $j_{1,1}\simeq3.831706$. Consequently,

\begin{equation} \label{IntervaloX1}
0\leq x<j_{1,1}.
\end{equation}

On the other hand, the condition $G_I(r)\geq0$ implies $1-2J_1(x)/x\geq0$, or equivalently $J_1(x)/x\leq1/2$. Since $\displaystyle\lim_{x\rightarrow0}J_1(x)/x=1/2$, this condition is also satisfied throughout the interval $0\leq x<j_{1,1}$, with $G_I(0)=0$. 
Using relation \eqref{RelacionXR}, the corresponding radial domain is
\begin{equation} \label{IntervaloR1}
0\leq r<
R_c\equiv
\frac{j_{1,1}^{\,2}}{4}A
\simeq3.67049\,A.
\end{equation}

Thus, within $0\leq r<R_c$, the running coupling satisfies $0\leq G_I(r)<G_0$, ensuring a real RG scale $k$.

\item At $r=R_c$, one has $G_I(r_c)=G_0$ and $k^2(r_c)=0$. For $r>R_c$, the Bessel function $J_1$ changes sign, leading to $G_I(r)>G_0$ and consequently to $k^2<0$, which would correspond to imaginary values of the RG scale $k$. Although the mathematical Bessel solution continues beyond this point and oscillates around $G_0$ with a decreasing amplitude proportional to $r^{-3/4}$, approaching $G_0$ as $r\to\infty$, such an extension is not physically admissible under our requirement $k^2>0$. For this reason, below we will restrict Region I to $R_I<R_c$, with the specific values of $R_I$ further constrained by the Israel junction conditions discussed below.

\item We also require the running coupling to be monotonically increasing throughout Region I, namely $G_I'(r)>0$. Differentiating Eq.~\eqref{Ratio1} and using Eq.~\eqref{RelacionXR}, we obtain

\begin{equation}
\label{DerivativeGI}
G_I'(r)
=
\frac{G_0}{r}\,J_2(x).
\end{equation}

Therefore, for $r>0$, the condition $G_I'(r)>0$ is equivalent to $J_2(x)>0$. The Bessel function of the first kind $J_2(x)$ of order two remains positive from the origin up to its first positive zero, denoted by $j_{2,1}$, where $J_2(j_{2,1})=0$ and $j_{2,1}\simeq5.135622$. Hence, the running coupling remains monotonically increasing for

\begin{equation}
\label{IntervaloX2}
0<x<j_{2,1} \,\, \Rightarrow \,\, 0<r<\frac{j_{2,1}^{\,2}}{4}\,A
\simeq6.59365\,A
\end{equation}

\item Since $j_{1,1}\simeq3.831706<j_{2,1}\simeq5.135622$, the interval already imposed in Eq.~\eqref{IntervaloX1} is more restrictive than the monotonicity condition \eqref{IntervaloX2}. Consequently, throughout the radial domain defined in Eq.~\eqref{IntervaloR1}, one simultaneously has $0\leq G_I(r)<G_0$ and $G_I'(r)>0$. 
\item Region I is defined by
\begin{equation}
0\leq r\leq R_I,
\end{equation}
where $R_I$ must lie within the radial domain \eqref{IntervaloR1} in order to satisfy the conditions described in the previous item.

\item On the other hand, as discussed below, the matching radius must satisfy $R_I<R_c$ in order to fulfill the Israel junction conditions at $r=R_I$.
    
\end{itemize}

\subsection{Israel Junction Conditions and Black Hole Geometry Without an Inner Horizon}

As explained above, our running gravitational coupling $G_I(r)$ should not be extrapolated to radial distances $r\geq R_I$ for the reasons discussed previously. Therefore, in this section, we investigate, through the Israel junction conditions, the possibility of describing the exterior region by a different running gravitational coupling $G_{II}(r)$. The resulting piecewise-defined running gravitational coupling is given by Eq.~\eqref{tramos}. A basic requirement is that, at low-energy and large-distance scales, the usual Newton coupling is recovered, $G_{II}(r)\to G_0$ for $r \to \infty$. Although several candidates satisfying this infrared behavior have been considered in the literature, in the present work we study the Bonanno running coupling \cite{Bonanno:2000ep,Falls:2010he}, given by
\begin{equation} \label{ExpresionG2Bonano}
G_{II}(r)
=
\frac{G_0r^3}
{
r^3+
\displaystyle\frac{\xi_{II}^{\,2}}{g_*}\,G_0
\left(r+\gamma mG_0\right)
}.
\end{equation}
It is worth emphasizing that the scale-identification mechanism adopted in Region II differs from the prescription introduced in Region I in our search for an integrable singularity near the radial origin. In the Bonanno--Reuter construction, the RG scale is identified according to $k_{II}(r)=\xi_{II}/D(r)$, where $\xi_{II}$ is a dimensionless parameter and $D(r)$ represents a characteristic length scale associated with the radial position, given by $D(r)=\sqrt{\frac{r^3}{r+\gamma G_0 m}}$. It is worth noting that $\gamma=9/2$ corresponds to the interpolation based on the radial proper distance in the Schwarzschild geometry, such that $D(r)\sim r$ at large radial distances. In particular, large values of $D(r)$ correspond to low-energy scales, so that $k_{II}(r)\rightarrow 0$ in the infrared regime. Since the scale-setting prescriptions employed in Regions I and II are different, there is no requirement to impose $\xi_{II}=\xi$. In the present work, however, we investigate the scenario in which the non-Gaussian ultraviolet fixed point $g_*$ of the dimensionless Newton coupling is the same in both regions.

Before testing the Israel junction conditions, we focus on the scenario in which the black-hole geometry possesses a single horizon. In this case, the metric function $f(r)$ undergoes only one sign change, from negative to positive as the radial coordinate increases. From Eq.~\eqref{GIOrigen}, we note that
\begin{equation} \label{fOrigen}
f(r)\big|_{r\rightarrow0}
\sim
1-\frac{g_*}{\xi}.
\end{equation}

Therefore, in order for the region near the radial origin to satisfy $f(r)<0$, we require
\begin{equation} \label{Condicion1Xi}
0<\xi<g_*.
\end{equation}

Together with the asymptotic behavior $f(r)\rightarrow1$ at large radial distances. In the following, we investigate whether the matched geometry can realize this scenario with only one such transition.

Due to the highly non-trivial and nonlinear character of the Israel junction conditions, particularly because of the presence of Bessel functions and their radial derivatives in the Region-I solution, we perform the matching numerically by appropriately adjusting the values of certain model parameters. For this purpose, we work in Planck units and set $G_0=1\,\ell_{\rm p}^{\,2}$. We choose $g_*=1$, which could be understood as a convenient order-unity reference value for the dimensionless ultraviolet fixed point, rather than as a unique prediction for its physical value. This choice could be particularly convenient for exploring the short-distance regime in which quantum-gravitational effects are expected to become relevant. From Eq.~\eqref{EcuacionA}, the characteristic radial scale then becomes $A=\xi\,m$. For simplicity, we choose this characteristic scale as our unit of length, $A=1$. Consequently, we use
\begin{equation} \label{mvsXi}
m=\frac{1}{\xi}.
\end{equation}

Furthermore, from the condition \eqref{Condicion1Xi}, the parameter $\xi$ must satisfy
\begin{equation}
\label{Condicion2Xi}
0<\xi<1.
\end{equation}

Furthermore, in accordance with the physical arguments discussed above, we adopt $\gamma=9/2$ for Region II. Thus, within this numerical setup, varying $\xi$ in the interval $0<\xi<1$ generates a one-parameter family of configurations with corresponding masses $m=1/\xi$. For each numerical choice of the ordered pair $\{\xi,m\}$, we impose the Israel junction conditions at the matching radius $r=R_I$ to determine the numerical values of the parameters $\{R_I,\xi_{II}\}$.

To construct a smooth geometry connecting Regions I and II, we impose the Israel--Darmois junction conditions \cite{Israel:1966rt} at the hypersurface $\Sigma$, identified with the matching radius $r=R_I$. These conditions require the continuity of both the first and second fundamental forms across $\Sigma$. The jump of any quantity $F$ across the matching surface will be denoted by
$[F]_{\Sigma}\equiv F\big|_{R_I^{,+}}-F\big|_{R_I^{\,-}}$. 

\begin{itemize}
    \item The first fundamental form requires continuity of the induced metric across $\Sigma$. For the line element given by equations \eqref{MetricEQC}, \eqref{fEQC}. This condition implies $[f]_{\Sigma}=0$, or equivalently $f_I(R_I)=f_{II}(R_I)$. Due that the same mass parameter $m$ is used on both sides of the matching surface, the first junction condition reduces to
\begin{equation}
\label{Israel1}
G_I(R_I)=G_{II}(R_I).
\end{equation}
\item \label{Condicion2JC} In this case, the continuity of the second fundamental form, through the Gauss--Codazzi equations, implies $[\mathcal{G}{\mu\nu}x^{\nu}]{\Sigma}=0$, where $x^{\nu}$ is the unit normal vector to the hypersurface $\Sigma$, directed along the radial direction. This condition implies
\begin{align}
\frac{R_I f_I'(R_I)+f_I(R_I)-1}{R_I^2}
&=
\frac{R_I f_{II}'(R_I)+f_{II}(R_I)-1}{R_I^2}, 
\nonumber\ \\ \label{Israel2}
\Rightarrow \quad G_I'(R_I)&=G_{II}'(R_I),  \end{align}
where we have used the first junction condition \eqref{Israel1}.
\end{itemize}

Under the numerical assumptions described above, for a fixed value of $\xi$, the first Israel junction condition, equation \eqref{Israel1}, gives
\begin{equation}
1-
\frac{J_1\!\left(2\sqrt{R_I}\right)}
{\sqrt{R_I}}
=
\frac{R_I^3}
{
R_I^3+
\xi_{II}^{\,2}
\left(
R_I+\frac{9}{2\xi}
\right)
}.
\end{equation}
meanwhile, the second Israel junction condition, Eq.~\eqref{Israel2}, gives

\begin{equation}
\frac{J_2\!\left(2\sqrt{R_I}\right)}
{R_I}
=
\frac{
\xi_{II}^{\,2}R_I^2
\left(
2R_I+\frac{27}{2\xi}
\right)
}{
\left[
R_I^3+
\xi_{II}^{\,2}
\left(
R_I+\frac{9}{2\xi_I}
\right)
\right]^2
}.
\end{equation}

In accordance with the assumptions described above, in Table~\ref{tab:matching} we display representative numerical configurations for $G_0=1$, $g_*=1$, $A=1$, and $\gamma=9/2$. Furthermore, from Eq.~\eqref{IntervaloR1}, we take $R_c=3.67049$. According to the condition in Eq.~\eqref{Condicion2Xi}, we consider values of the Region-I parameter in the range $0.1991\leq \xi_I\leq0.99$, from which the corresponding values of the mass parameter $m$ are directly obtained through Eq.~\eqref{mvsXi}. Then, for each ordered pair $\{\xi_I,m\}$, the Israel junction conditions in Eqs.~\eqref{Israel1} and \eqref{Israel2} determine the matching parameters ${R_I,\xi_{II}}$. As can be seen from the fifth column of Table~\ref{tab:matching}, the ratio $R_I/R_c$ remains smaller than unity over the entire range considered, and therefore the condition $R_I<R_c$ established in Eq.~\eqref{IntervaloR1} is satisfied in all cases. Interestingly, although the mass parameter increases significantly across the considered configurations, the matching radius varies only mildly, remaining in the narrow interval $R_I\simeq2.19$--$2.31$, corresponding to $R_I/R_c\simeq0.60$--$0.63$. Thus, the matching surface remains well inside the radial domain of validity of Region I throughout the numerical family.

Once the space of parameters $\{\xi_I,m,R_I,\xi_{II}\}$ have been determined, we proceed to study the event-horizon structure of the resulting geometry. In accordance with the main purpose of this work, we are particularly interested in configurations possessing a single event horizon, with no additional inner horizon that could potentially be associated with an unstable Cauchy horizon. From Eq.~\eqref{fEQC}, the location of an event horizon is determined by the condition $f(r_h)=0$, which can be written as
\begin{equation}
\label{Rh}
r_h=2\,m\,G(r_h),
\end{equation}
where the running gravitational coupling $G(r)$ is defined piecewise according to Eq.~\eqref{tramos}. Since the Region-I solution involves Bessel functions and the matching parameters are themselves determined numerically, we study the horizon structure numerically over the interval $0.1991\leq\xi_I\leq0.99$. For each value of $\xi_I$, the corresponding values of $m$, $R_I$, and $\xi_{II}$ are first obtained from the relations discussed above and from the Israel junction conditions.

\begin{figure}[h]
  \begin{center}
      \includegraphics[width=5in]{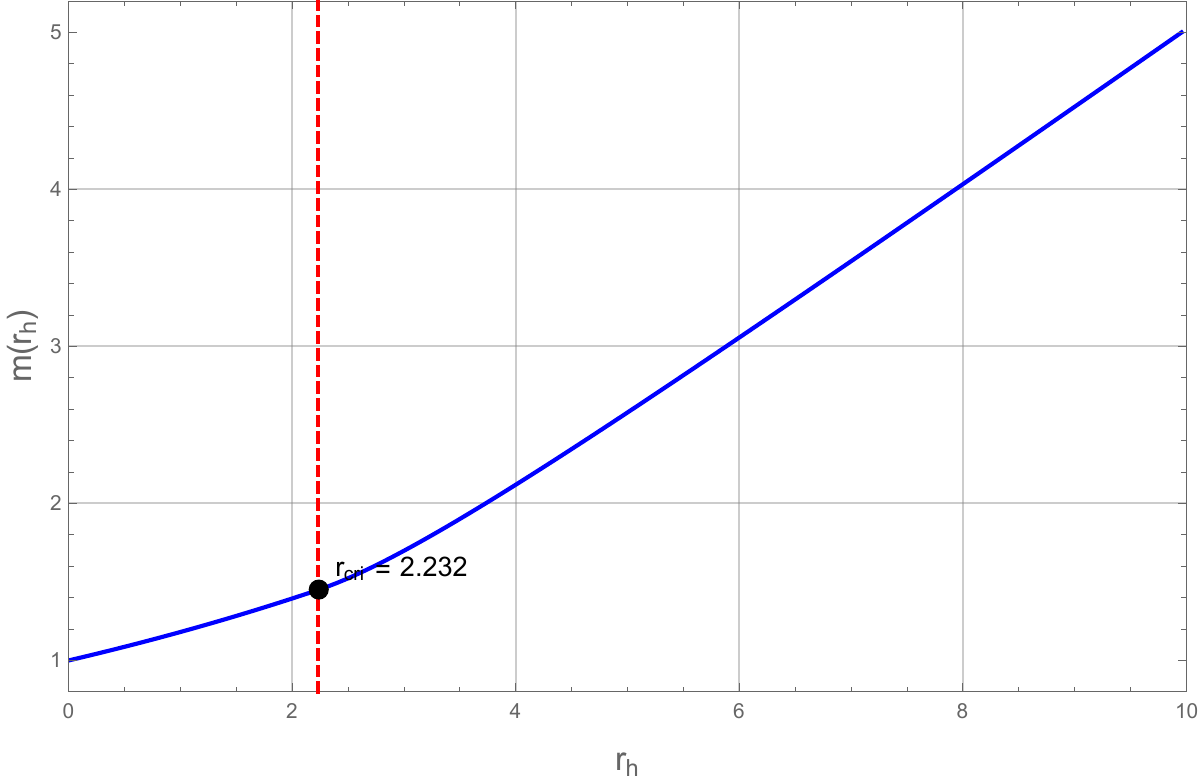}
            \caption{Mass parameter $m(r_h)$ versus the event horizon radius $r_h$, for $G_0=1$, $g_*=1$, $A=1$, and $\gamma=9/2$}
\label{FigMrh}
  \end{center}
\end{figure}

In order to determine whether Eq.~\eqref{Rh} must be solved using $G_I$ or $G_{II}$, we evaluate the metric function at the matching surface. The numerical procedure can be summarized as follows:
\begin{itemize}
\item From Eqs.~\eqref{fOrigen}, \eqref{Condicion1Xi}, and \eqref{Condicion2Xi}, we have $f(r\rightarrow0)<0$.Therefore, the region sufficiently close to the radial origin belongs to the trapped sector of the geometry.
\item If $f(R_I)>0$, the metric function must have changed sign before reaching the matching surface $r=R_I$. Consequently, $r_h<R_I$ and the event horizon lies in Region I. Its location is therefore determined by substituting $G(r)=G_I(r)$ into Eq.\eqref{Rh}.

In Table \ref{tab:matching}, the last two columns indicate the event horizon radius $r_h$ and the ratio $r_h/R_I$, respectively. This ratio is smaller than unity because the event horizon lies within Region I. In Fig. \ref{FigMrh}, we display $r_h$ versus the mass parameter $m(r_h)$. We observe that, for each value of the mass parameter, there is a unique event horizon radius, indicating the absence of degeneracy and, consequently, the absence of an inner horizon. The event horizon radii located to the left of the red dashed line correspond to configurations satisfying $r_h/R_I<1$.
\item If $f(R_I)=0$, the event horizon coincides with the matching surface, $r_h=R_I\equiv r_{\rm cri}$. Its location is therefore determined by substituting either $G(r)=G_I(r)$ or $G(r)=G_{II}(r)$ into Eq.~\eqref{Rh}. This configuration defines the critical point separating the two branches of the horizon curve. In Table \ref{tab:matching}, this can be observed in the fifth row, where $r_h/R_I=1$. In Fig. \ref{FigMrh}, it corresponds to the red dashed line.
\item If $f(R_I)<0$, the geometry remains trapped when the matching surface is crossed. Since $G_{II}(r)\rightarrow G_0$ and consequently $f_{II}(r)\rightarrow 1$ as $r\rightarrow\infty$, the zero of the metric function occurs in Region II. Hence, $r_h>R_I$, and its position is obtained by substituting $G(r)=G_{II}(r)$ into Eq.~\eqref{Rh}. In Table~\ref{tab:matching}, the ratio $r_h/R_I$ is greater than unity because the event horizon lies within Region II. In Fig.\ref{FigMrh}, we observe that, for each value of the mass parameter, there is a unique event horizon radius, indicating the absence of degeneracy and, consequently, the absence of an inner horizon. The event horizon radii located to the right of the red dashed line correspond to configurations satisfying $r_h/R_I>1$.

\item Repeating this procedure continuously over the considered range of $\xi_I$ generates the relation between the event horizon radius $r_h$ and the mass parameter $m$, shown in Fig. \ref{FigMrh}.
\end{itemize}

\begin{table}[ht]
\centering
\caption{Representative numerical configurations obtained from the matching conditions for $G_0=1$, $g_*=1$, $A=1$, $\gamma=9/2$, and $R_c=3.67049$.}
\begin{tabular}{ccccccc}
\hline\hline
$\xi$ &
$m$ &
$\xi_{II}$ &
$R_I$ &
$R_I/R_c$ &
$r_h$ &
$r_h/R_I$
\\
\hline
0.9900 & 1.0101 & 0.7001 & 2.1943 & 0.5978 & 0.0603 & 0.0275 \\
0.9000 & 1.1111 & 0.6777 & 2.2050 & 0.6007 & 0.6323 & 0.2867 \\
0.8000 & 1.2500 & 0.6501 & 2.2177 & 0.6042 & 1.3401 & 0.6043 \\
0.7170 & 1.3947 & 0.6247 & 2.2287 & 0.6072 & 2.0000 & 0.8974 \\
0.6900 & 1.4494 & 0.6159 & 2.2325 & 0.6082 & 2.2325 & 1.0000 \\
0.4718 & 2.1194 & 0.5313 & 2.2651 & 0.6171 & 4.0000 & 1.7659 \\
0.3271 & 3.0571 & 0.4560 & 2.2896 & 0.6238 & 6.0000 & 2.6206 \\
0.2479 & 4.0334 & 0.4041 & 2.3041 & 0.6277 & 8.0000 & 3.4720 \\
0.1991 & 5.0219 & 0.3662 & 2.3136 & 0.6303 & 10.0000 & 4.3224 \\
\hline\hline
\end{tabular}
\label{tab:matching}
\end{table}

An interesting feature is that the critical point does not signal the appearance of an additional horizon. Rather, it marks the transition between the two radial descriptions of the same event horizon: below $r_{\rm cri}$, the horizon is contained in the quantum-corrected inner region, whereas above $r_{\rm cri}$, it is located in the exterior RG-improved geometry. Within the entire parameter range investigated numerically, we find a single zero of the metric function for each value of the mass parameter. Therefore, the numerical solutions realize precisely the scenario sought in this work: a black-hole geometry with a single event horizon and without an additional potentially unstable inner horizon.

Moreover, since the numerical analysis is performed with $G_0=1$ and $A=1$, the transition between the two descriptions takes place at radial scales of order unity in Planck units. This is consistent with the interpretation that the modification of the gravitational coupling becomes relevant in the short-distance regime, while the exterior geometry progressively approaches the usual infrared behavior.

\begin{figure}[h]
  \begin{center}
      \includegraphics[width=7in]{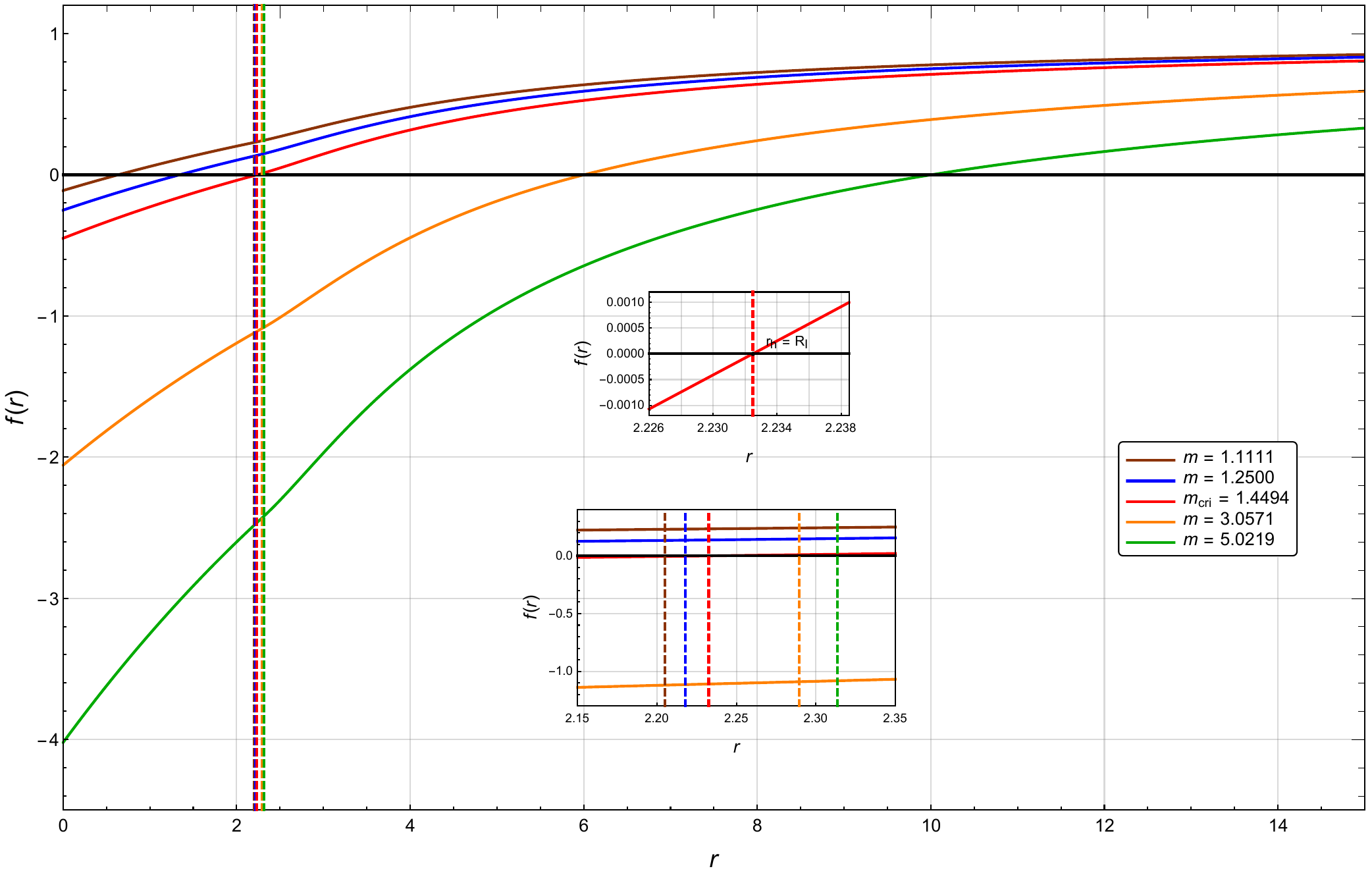}
            \caption{Metric function $f(r)$ versus the radial coordinate $r$, for $G_0=1$, $g_*=1$, $A=1$, and $\gamma=9/2$, considering different values of the mass parameter $m=1.1111, 1.25, m_{\rm cri}=1.4494, 3.0571$, and $5.0219$. The values of $R_I$ are indicated by the vertical dashed lines.}
\label{Figf(r)}
  \end{center}
\end{figure}

Figure~\ref{Figf(r)} displays the behavior of the metric function $f(r)$ for different values of the mass parameter $m$. The values of $R_I$ corresponding to the different values of $m$ are indicated by the vertical dashed lines. In the lower zoom of Fig.~\ref{Figf(r)}, we observe that the value of the boundary of Region I, $R_I$, increases as the mass parameter $m$ increases. We observe the existence of a critical mass $m_{\rm cri}$ for which $r_h=R_I$, that is, the boundary of Region I coincides with the location of the event horizon. This feature can be clearly seen in the upper zoom of Fig.~\ref{Figf(r)}.  Radial values satisfying $r<R_I$ belong to Region I and are characterized by the running gravitational coupling $G_I(r)$ discussed above, whereas radial values satisfying $r>R_I$ belong to Region II and are characterized by the exterior running coupling $G_{II}(r)$. For masses larger than the critical value, $m>m_{\rm cri}$, the event horizon is located in Region II, while for $m<m_{\rm cri}$ it lies inside Region I. At the critical configuration, the horizon coincides precisely with the matching surface. Remarkably, throughout the range of masses considered here, the metric function exhibits only one zero. Thus, within the numerical domain explored, the resulting geometry possesses a single event horizon and no additional potentially unstable inner horizon of the type that commonly appears in regular black-hole geometries with a dS core. The configurations with $m<m_{\rm cri}$ could be particularly interesting from the viewpoint of short-distance quantum-gravity effects, since in this regime the event horizon itself is located inside Region I, where the running coupling $G_I(r)$ is governed by the high-energy scale identification introduced above. These configurations could therefore describe a genuinely Planckian regime in which short length scales and high RG scales are directly associated with the formation of the integrable singularity. In this sense, the transition at $m=m_{\rm cri}$ could separate configurations whose horizon already lies within the quantum-corrected inner region from those for which the horizon is located in the exterior RG-improved geometry.

\section{Effective Cored Density from an RG-Improved NFW Profile}

As discussed in the Introduction, BH geometries embedded in NFW galactic halos have received increasing attention in recent years; see, for example, Ref. \cite{Xu:2018wow} and subsequent applications. Here, we extend our gravitational construction by incorporating an additional DM source described by the Navarro--Frenk--White (NFW) profile \cite{Navarro:1995iw}, 
\begin{equation}
\rho_{\rm NFW}(r)=
\frac{\rho_s}{
\displaystyle \frac{r}{r_s}
\left(1+\frac{r}{r_s}\right)^2
},
\end{equation}
where $\rho_s$ is the characteristic density and $r_s$ is the characteristic scale radius of the dark matter halo. At short radial distances, $r\ll r_s$, this profile behaves as $\rho_{\rm NFW}(r)\sim r^{-1}$, exhibiting a divergent density at the radial origin. Motivated by the physical implications of such a divergence, as discussed in the Introduction, in this section we investigate whether the running gravitational coupling associated with our quantum-gravitational construction can soften this behavior and lead to a finite effective dark matter density near the origin. In particular, since $G_I(r)\sim r$ in the UV regime, we develop a procedure in which the effective density weighted by the gravitational coupling approaches a finite constant as $r\to0$, despite the divergent behavior of the original NFW profile. This provides a physical motivation for incorporating the dark matter distribution through the same scale-dependent gravitational coupling characterizing the seed geometry. To this end, we retain the same renormalization-group scale identification $k(r)$ given in Eq.~\eqref{IdentificacionMia}, together with the corresponding effective gravitational coupling $G(r)$ previously determined in each region, as described by Eqs.~\eqref{tramos}, \eqref{G1Expresion}, and \eqref{ExpresionG2Bonano}. In doing so, we maintain the physical assumptions previously established regarding the radial domains of validity of each region. This choice is based on the assumption that the scale identification characterizes the corrections associated with the quantum-gravitational sector of the seed geometry and does not directly depend on the particular nature of additional matter sources. In this sense, we assume that the effective gravitational coupling has a universal character, such that different sources, regardless of their composition or physical nature, interact gravitationally through the same $G(r)$. Within this phenomenological procedure, the incorporation of dark matter modifies the geometry through an additional contribution to the energy-momentum tensor, without necessarily requiring a redefinition of the previously established scale identification. In this way, we can investigate the effects of dark matter on the geometric structure and horizon configuration while preserving the UV behavior of the gravitational coupling associated with the seed space-time.

We begin by rewriting the effective Einstein equations \eqref{EinsteinEQC} as
\begin{equation} \label{MultiplesFuentes}
\mathcal{G}^{\mu}{}_{\nu}
=
8\pi G(r)\left(T^\mu{}_\nu\right)_{\rm qc}
+
8\pi\,\alpha \,G_0\,\left(T^\mu{}_\nu\right)_{\rm eff-\alpha},
\end{equation}
where $G(r)$ is the same running gravitational coupling
established in the previous sections, as described by
Eqs.~\eqref{tramos}, \eqref{G1Expresion}, and
\eqref{ExpresionG2Bonano}. The tensor $\left(T^\mu{}_\nu\right)_{\rm qc}$ corresponds to the effective energy-momentum tensor associated with the quantum contribution, whose structure was established in the previous sections, while $\alpha$ is a parameter controlling the contribution of the additional matter source, and
$\left(T^\mu{}_\nu\right)_{\rm eff-\alpha}$ represents the effective energy-momentum tensor associated with this additional matter source, whose energy density is modified by the running gravitational coupling according to our scale identification. 
Accordingly, we write
\begin{equation}
\left(T^\mu{}_\nu\right)_{\rm eff-\alpha}
=
\operatorname{diag}
\left(
-\rho_{\rm eff-\alpha},
-\rho_{\rm eff-\alpha},
p_{\rm eff-\alpha},
p_{\rm eff-\alpha}
\right),
\end{equation}

Thus, the effective energy
density and tangential pressure are given by
\begin{equation}
\rho_{\rm eff-\alpha}(r)
=
\frac{G(r)}{G_0}\rho_{\rm NFW}(r),
\qquad
p_{\rm eff-\alpha}(r)
=
\frac{G(r)}{G_0}p(r).
\end{equation}
Here, $\rho_{\rm NFW}(r)$ denotes the Navarro--Frenk--White
dark matter density profile, while $p(r)$ represents
the tangential pressure associated with the additional
matter source before incorporating the running
gravitational coupling.  From Eq.~\eqref{GIOrigen}, we can observe that, at short radial distances, the effective  energy density behaves as
\begin{equation}
\rho_{\rm eff-\alpha}(r) \big |_{r \to 0}
\sim
\frac{\rho_s r_s}{2A}.
\end{equation}
Therefore, the effective energy density approaches a finite constant near the radial origin, despite the divergent behavior of the original NFW profile. This result is consistent with what we seek to achieve in this section, namely, to obtain a finite central dark matter density through the ultraviolet behavior of the running gravitational coupling.

Following a procedure analogous to gravitational decoupling \cite{Ovalle:2017fgl}, Eq.~\eqref{MultiplesFuentes} can be understood in terms of the decoupling of gravitational sources. Following this algorithm, we consider the following static and spherically symmetric geometry:
\begin{equation} \label{MasaMGD}
ds^2
=
-f_{\rm mgd}(r)\,dt^2
+
\frac{dr^2}{f_{\rm mgd}(r)}
+
r^2d\Omega^2,
\end{equation}
where the metric function $f_{\rm mgd}(r)$ can be interpreted as a minimal geometric deformation (MGD) of the seed geometry, such that
\begin{equation}
\label{DeformacionMGD}
f_{\rm mgd}(r)
=
f(r)-\alpha\,\frac{2m_\alpha(r)}{r}=
1-\frac{2mG(r)}{r}
-\alpha\,\frac{2 m_\alpha(r)}{r}.
\end{equation}

Here, $m_\alpha(r)$ characterizes the geometric deformation induced by the additional matter source, while $\alpha$ controls its contribution to the resulting geometry. We consider as our seed geometry the spacetime characterized by the metric function $f(r)$ given in Eq.~\eqref{fEQC}, where the running gravitational coupling is described by Eqs.~\eqref{tramos}, \eqref{G1Expresion}, and \eqref{ExpresionG2Bonano}. The effective matter sources associated with the seed geometry are those established in the previous sections, particularly in Eqs.~\eqref{TEQC} and \eqref{rhoEQC}. 
In this way, the effective Einstein equations can be split into two sets. 

The first set corresponds to the seed equations of order $\alpha^0$, which reproduce the original geometry characterized by $f(r)$ and its associated effective quantum-gravitational sources:
\begin{equation}
\mathcal{G}^{\mu}{}_{\nu}
=
8\pi G(r)
\left(T^\mu{}_\nu\right)_{\rm qc}.
\end{equation}

The second set corresponds to the quasi-Einstein--like equations associated with the additional matter source, obtained by collecting the terms proportional to $\alpha$. For the geometric deformation introduced above, these equations are given by
\begin{equation}
\label{QuasiEinstein1}
\frac{2m_\alpha'(r)}{r^2}
=
8\pi G_0\rho_{\rm eff-\alpha}(r),
\end{equation}
and
\begin{equation}
\label{QuasiEinstein2}
\frac{m_\alpha''(r)}{r}
=
-8\pi G_0p_{\rm eff-\alpha}(r).
\end{equation}

Using the previously established definitions of the effective energy density and tangential pressure, these equations become
\begin{equation}
m_\alpha'(r)
=
4\pi G_0r^2\rho_{\rm eff-\alpha}(r)
=
4\pi r^2G(r)\rho_{\rm NFW}(r),
\end{equation}
and
\begin{equation}
p_{\rm eff-\alpha}(r)
=
-\frac{m_\alpha''(r)}{8\pi G_0r}
=
\frac{G(r)}{G_0}p(r).
\end{equation}

It is worth emphasizing that the gravitational decoupling procedure allows us to identify two separately conserved effective sectors. To verify this property, we define
\begin{equation}
\left(T^\mu{}_\nu\right)_{\rm seed}
=
G(r)\left(T^\mu{}_\nu\right)_{\rm qc}.
\end{equation}
The total effective energy-momentum tensor can then be written as
\begin{equation}
\left(T^\mu{}_\nu\right)_{\rm total}
=
\left(T^\mu{}_\nu\right)_{\rm seed}
+
\alpha G_0
\left(T^\mu{}_\nu\right)_{\rm eff-\alpha}.
\end{equation}

According to the contracted Bianchi identities, the total effective energy-momentum tensor satisfies
\begin{equation}
\nabla_\mu
\left(T^\mu{}_\nu\right)_{\rm total}
=0,
\end{equation}
where $\nabla_\mu$ denotes the covariant derivative associated with the full metric $f_{\rm mgd}(r)$. Using the effective density and pressure components established for the seed geometry, together with Eqs.~\eqref{QuasiEinstein1} and \eqref{QuasiEinstein2} for the additional source, we can verify that both sectors independently satisfy this conservation equation. Therefore,
\begin{equation}
\label{ConservationSeed}
\nabla_\mu
\left(T^\mu{}_\nu\right)_{\rm seed}
=0,
\end{equation}
and
\begin{equation}
\label{ConservationAlpha}
\nabla_\mu
\left(T^\mu{}_\nu\right)_{\rm eff-\alpha}
=0.
\end{equation}

Thus, within our gravitational decoupling construction, the seed and additional matter sectors are separately conserved. Although both sectors contribute to the same spacetime geometry and therefore remain gravitationally coupled, there is no direct exchange of energy and momentum between them. This property is compatible with the phenomenological description of a dark matter sector that interacts gravitationally with the seed geometry, without introducing additional non-gravitational interactions between the two effective sources.

Integrating the equation for the radial mass function, we obtain
\begin{equation}
\label{MasaAlpha}
m_\alpha(r)
=
4\pi\int r^2G(r)\rho_{\rm NFW}(r)\,dr+K,
\end{equation}
where $K$ is an integration constant.

Before investigating the geometric properties of the resulting spacetime, we must verify that the Israel--Darmois junction conditions remain satisfied at the matching surface $r=R_I$. Although the running gravitational coupling $G(r)$ was previously matched across Regions I and II, the additional matter source introduces a geometric deformation through the mass function defined in Eq.~\eqref{MasaAlpha}. Consequently, the continuity of $G(r)$ and its first derivative does not necessarily guarantee the continuity of the deformed geometry. Therefore, the junction conditions must be examined to ensure a consistent matching between both regions without introducing an additional thin shell at the matching surface.

\subsection{Israel--Darmois Junction Conditions in the Presence of Dark Matter}

We now examine the Israel--Darmois junction conditions for the modified geometry given in Eq.~\eqref{MasaMGD}, taking into account the additional mass contribution defined in Eq.~\eqref{MasaAlpha}. Since we retain the same scale identification and running gravitational coupling established in the previous sections, the matching conditions for the seed geometry remain satisfied at $r=R_I$, namely,
\begin{equation}
\label{IsraelMGD1}
G_I(R_I)=G_{II}(R_I),
\qquad
G_I'(R_I)=G_{II}'(R_I).
\end{equation}

These conditions were previously established in Eqs.~\eqref{Israel1} and \eqref{Israel2}. However, the additional matter source introduces a geometric deformation through $m_\alpha(r)$, whose continuity must also be examined.

The continuity of the first fundamental form requires
\begin{equation}
\label{IsraelMGD2}
f_{{\rm mgd},I}(R_I)
=
f_{{\rm mgd},II}(R_I).
\end{equation}
Using Eq.~\eqref{DeformacionMGD} and the first condition in Eq.~\eqref{IsraelMGD1}, we obtain
\begin{equation}
\label{IsraelMasa1}
m_{\alpha,I}(R_I)
=
m_{\alpha,II}(R_I).
\end{equation}

On the other hand, the continuity of the second fundamental form requires
\begin{equation}
\label{IsraelMGD3}
f_{{\rm mgd},I}'(R_I)
=
f_{{\rm mgd},II}'(R_I).
\end{equation}
Taking into account Eqs.~\eqref{IsraelMGD1} and \eqref{IsraelMasa1}, this condition reduces to
\begin{equation}
\label{IsraelMasa2}
m_{\alpha,I}'(R_I)
=
m_{\alpha,II}'(R_I).
\end{equation}

From Eq.~\eqref{MasaAlpha}, we have
\begin{equation}
m_{\alpha,i}'(r)
=
4\pi r^2G_i(r)\rho_{\rm NFW}(r),
\qquad i=I,II.
\end{equation}
Since the same NFW density profile is considered in both regions and $G_I(R_I)=G_{II}(R_I)$, we obtain
\begin{equation}
m_{\alpha,I}'(R_I)
=
m_{\alpha,II}'(R_I).
\end{equation}
Therefore, the second junction condition is automatically satisfied once the continuity of the mass function is imposed.

Finally, writing
\begin{equation}
m_{\alpha,i}(r)
=
4\pi\int^r s^2G_i(s)\rho_{\rm NFW}(s)\,ds
+K_i,
\end{equation}
the first junction condition determines the relation between the integration constants in Regions I and II:
\begin{equation}
\label{IsraelConstantesMGD}
K_{II}-K_I
=
4\pi\left[
\int^{R_I}s^2G_I(s)\rho_{\rm NFW}(s)\,ds
-
\int^{R_I}s^2G_{II}(s)\rho_{\rm NFW}(s)\,ds
\right],
\end{equation}
where the integration limits are understood to be fixed consistently in each primitive.

Consequently, provided that the integration constants can be chosen to satisfy Eq.~\eqref{IsraelConstantesMGD}, the additional matter source does not require a modification of the previously determined matching radius $R_I$ or the parameters characterizing the running gravitational coupling in either region. Thus, the Israel--Darmois junction conditions remain satisfied for the modified geometry, without introducing an additional thin shell at the matching surface.

\subsection{Black Hole Geometry}

We now proceed to determine the explicit form of the mass function $m_{\alpha,I}(r)$ associated with the additional matter source in Region I. To this end, we substitute the running gravitational coupling $G_I(r)$ given in Eq.~\eqref{G1Expresion}, together with the NFW density profile, into Eq.~\eqref{MasaAlpha}. Since the resulting integral involves Bessel functions, we perform an integration by parts and subsequently express the remaining contribution as a convergent series involving Gauss hypergeometric functions.

Accordingly, the mass function in Region I is given by
\begin{equation}
m_{\alpha,I}(r)
=
4\pi G_0\rho_s r_s^3
\int
\frac{r}{(r+r_s)^2}
\left[
1-\sqrt{\frac{A}{r}}
J_1\left(2\sqrt{\frac{r}{A}}\right)
\right]dr+K_I,
\end{equation}
where $K_I$ is an integration constant associated with Region I. Introducing the dimensionless variable
\begin{equation}
x=\sqrt{\frac{r}{r_s}},
\end{equation}
the mass function can be expressed as
\begin{equation}
\label{MasaAlphaI}
m_{\alpha,I}(r)
=
4\pi\left[
I_1(x)-I_{2A}(x)+I_{2B}(x)
\right]+K_I,
\end{equation}
where
\begin{equation}
\label{IntegralI1}
I_1(x)
=
G_0\rho_s r_s^3
\left[
\ln(1+x^2)
-\frac{x^2}{1+x^2}
\right],
\end{equation}
\begin{equation}
\label{IntegralI2A}
\begin{aligned}
I_{2A}(x)
={}&G_0\rho_s r_s^3
\sqrt{\frac{A}{r_s}}
J_1\left(2x\sqrt{\frac{r_s}{A}}\right)\times
\left[
\arctan x-\frac{x}{1+x^2}
\right],
\end{aligned}
\end{equation}
and
\begin{equation}
\label{IntegralI2B}
\begin{aligned}
I_{2B}(x)
={}&G_0\rho_s r_s^3
\sum_{n=0}^{\infty}
\frac{(-1)^n}{n!(n+1)!}
\left(\frac{r_s}{A}\right)^n \times
\left[
x^{2n+1}\arctan x
-x^{2n+2}
{}_2F_1\left(
1,n+1;n+2;-x^2
\right)
\right].
\end{aligned}
\end{equation}

Here, ${}_2F_1$ denotes the Gauss hypergeometric function. The infinite series provides an exact analytical representation of the corresponding integral and converges absolutely for any finite $x\geq0$, assuming $A>0$ and $r_s>0$. Therefore, Eq.~\eqref{MasaAlphaI}, together with Eqs.~\eqref{IntegralI1}, \eqref{IntegralI2A}, and \eqref{IntegralI2B}, determines the mass function associated with the additional matter source in Region I. It is straightforward to verify that, at short length scales,
the mass function $m_{\alpha,I}(r)$ behaves as
\begin{equation} \label{MasaAlpha1Origen}
m_{\alpha,I}(x)
\sim
K_I+
4\pi G_0\rho_s r_s^3
\left[
\frac{r_s}{6A}x^6+O(x^8)
\right] \sim K_I+
\frac{2\pi G_0\rho_s r_s}{3A}r^3
+O(r^4).
\end{equation}

In order for the mass function $m_{\alpha,I}(r)$ to vanish
in the short-distance limit $r\to0$, we choose $K_1=0$. Taking into account Eqs.~\eqref{fOrigen} and \eqref{MasaAlpha1Origen}, together with the geometry \eqref{DeformacionMGD}, the metric function behaves at short length scales as
\begin{equation}
\left.
f_{\rm mgd}(r)
\right|_{r\to0}
\sim
1-\frac{g_*}{\xi}
-\alpha\,
\frac{4\pi G_0\rho_s r_s}{3A}\,r^2.
\end{equation}
We note that, for values of $g_*/\xi$ sufficiently larger than unity, the second term becomes dominant at short length scales compared with the third term, which is of order $r^2$. Thus,
\begin{equation} \label{fMGDorigen}
\left.
f_{\rm mgd}(r)
\right|_{r\to0}
\sim
1-\frac{g_*}{\xi}.
\end{equation}
Therefore, the presence of the effective NFW source does not modify the leading short-distance sign of the metric function, and one still has $f_{\rm mgd}(r\to0)<0$.

We now proceed to determine the mass function associated with the additional matter source in Region II. To this end, we substitute the Bonanno--Reuter running gravitational coupling given in Eq.~\eqref{ExpresionG2Bonano}, together with the NFW density profile, into Eq.~\eqref{MasaAlpha}. Accordingly, we obtain
\begin{equation}
\label{MasaAlphaIIIntegral}
\begin{aligned}
m_{\alpha,II}(r)
={}&4\pi G_0\rho_s r_s^3
\int
\frac{r^4\,dr}
{(r+r_s)^2
\left[
r^3+
\displaystyle\frac{\xi_{II}^{\,2}}{g_*}G_0
(r+\gamma mG_0)
\right]}
+K_{II},
\end{aligned}
\end{equation}
where $K_{II}$ is the integration constant associated with Region II. Although the integral in Eq.~\eqref{MasaAlphaIIIntegral} admits an exact analytical solution in terms of elementary functions, its explicit expression is rather cumbersome. In particular, its evaluation requires a partial fraction decomposition involving the real roots of the cubic polynomial appearing in the denominator, together with the determination of the corresponding coefficients. For the positive parameters considered in our model, this polynomial possesses a unique negative real root, while the remaining roots form a complex-conjugate pair. The resulting analytical expression can be written in terms of logarithmic and inverse trigonometric functions. For completeness, the mathematical details and the corresponding analytical solution are presented in Appendix~\ref{AppendixMassII}.

Nevertheless, following the numerical strategy adopted in the previous section, and given the highly non-trivial character of the resulting metric function and the corresponding horizon equations, we investigate the black-hole geometry numerically. This procedure allows us to examine the horizon structure and the effects of the additional dark matter source while consistently incorporating the matching conditions between Regions I and II. For the numerical analysis, we retain the RG scale-setting prescription and the piecewise running gravitational coupling adopted for the seed geometry. We use the values of $G_0$, $g_*$, $A$, and $\gamma$ considered in the preceding section, together with the same values of $\xi$, with the corresponding masses determined by Eq.~\eqref{mvsXi}. In addition, consistently with the short-distance behavior discussed above, we take $K_I=0$. Since Eq.~\eqref{IsraelMGD1} reproduces the continuity conditions given by Eqs.~\eqref{Israel1} and \eqref{Israel2}, the values of $R_I$ and $\xi_{II}$ for each pair ${\xi,m}$ coincide with those obtained in the previous section. The additional matter sector introduces an additional matching condition for the geometry, with $K_{II}$ fixed by Eq.~\eqref{IsraelConstantesMGD}. For each configuration $\{\xi,m,R_I,\xi_{II}\}$, the resulting value of $K_{II}$ ensures the continuity of $m_\alpha(r)$ and, consequently, of the full MGD geometry across the matching surface. Therefore, the numerical procedure follows the same strategy as for the seed geometry, with the corresponding modifications induced by the terms of order $\alpha$. Table~\ref{tab:matchingNFW} summarizes representative configurations obtained following this procedure. The first five columns of Table~\ref{tab:matchingNFW} reproduce the same matching family obtained for the seed geometry in Table~\ref{tab:matching}. The sixth column gives the value of $K_{II}$ required by Eq.~\eqref{IsraelConstantesMGD}. 

However, our RG energy-scale identification, together with the inclusion of additional matter sources of order $\alpha$, which in turn deform the geometry proportionally to the control parameter $\alpha$, modifies the horizon structure. The numerical identification of the radial region containing the event horizon follows the same logic as in the seed geometry, but now using the complete metric function $f_{\rm mgd}(r)$. From Eq.~\eqref{fMGDorigen}, the metric function remains negative sufficiently close to the radial origin, $f_{\rm mgd}(r\to0)<0$. Therefore, if $f_{\rm mgd}(R_I)>0$, the metric function must cross zero before reaching the matching surface, and the event horizon lies in Region I, $r_h<R_I$. In this case, Eq.
\begin{equation}
\label{RhMGD}
f_{\rm mgd}(r=r_h)
=0=1-\frac{2mG(r_h)}{r_h}
-\alpha\,\frac{2 m_\alpha(r_h)}{r_h}.
\end{equation}
is solved using $G_I(r)$. If $f_{\rm mgd}(R_I)=0$, the horizon coincides with the matching surface, $r_h=R_I\equiv r_{\rm cri}$. Finally, if $f_{\rm mgd}(R_I)<0$, the geometry remains trapped when crossing into Region II, and the event horizon lies at $r_h>R_I$, where Eq.~\eqref{RhMGD} must be solved using $G_{II}(r)$.

The corresponding event-horizon radii and the ratios $r_h/R_I$ are displayed in the seventh and eighth columns of Table~\ref{tab:matchingNFW}, respectively. A relevant effect of the additional matter source is the displacement of the event horizon toward larger radial values for the configurations considered here. As a consequence, the critical configuration is also shifted. In the absence of the additional NFW sector, the critical configuration found in Table~\ref{tab:matching} corresponds to $m_{\rm cri}\simeq1.4494$ and $r_{\rm cri}=R_I\simeq2.2325$, whereas in the MGD geometry Table~\ref{tab:matchingNFW} gives $m_{\rm cri}\simeq1.3745$ and $r_{\rm cri}=R_I\simeq2.2273$. Thus, the additional matter contribution causes the transition $r_h=R_I$ to occur at a smaller value of the mass parameter.

\begin{figure}[h]
  \begin{center}
      \includegraphics[width=5in]{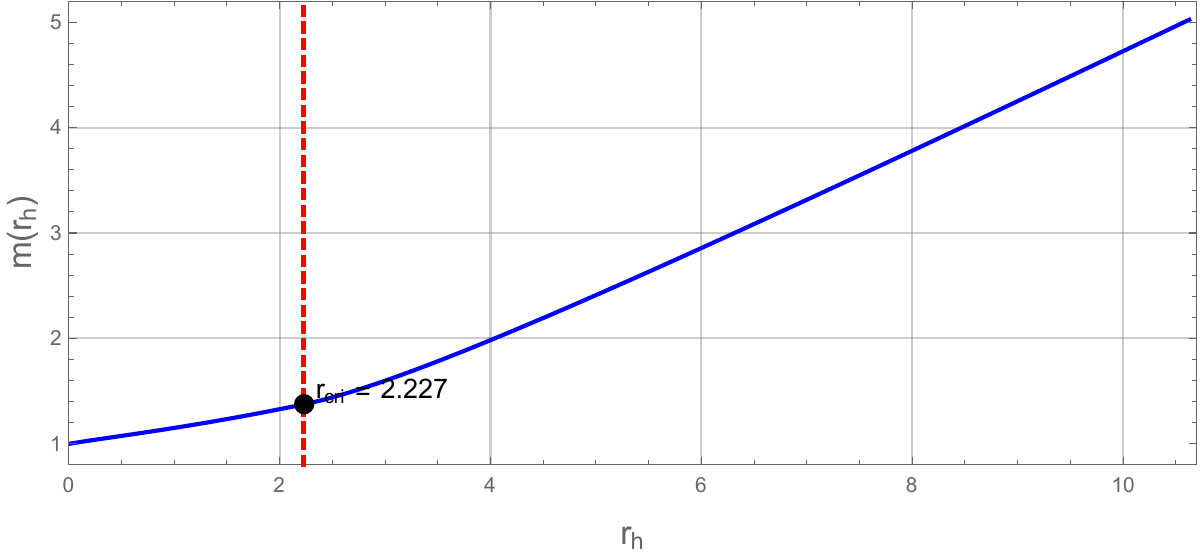}
            \caption{Mass parameter $m(r_h)$ versus the event horizon radius $r_h$, for $G_0=1$, $g_*=1$, $A=1$, $\gamma=9/2$ $\alpha=1$, $\rho_s=0.02$, $r_s=1$, and $K_I=0$}
\label{FigMrhMGD}
  \end{center}
\end{figure}

Figure~\ref{FigMrhMGD} displays the resulting relation between the mass parameter and the event-horizon radius. The red dashed line identifies the critical radius $r_{\rm cri}\simeq2.2273$. Configurations lying to the left of this line satisfy $r_h<R_I$ and therefore possess an event horizon located inside Region I, whereas configurations to the right satisfy $r_h>R_I$, with the event horizon located in Region II. The critical configuration separates these two regimes and satisfies $r_h=R_I$. As in the seed geometry, the relation remains single valued over the numerical domain considered: for each value of the mass parameter there is a unique event-horizon radius. Remarkably, this yields a black-hole geometry without an additional inner horizon associated with a potentially unstable de Sitter core.

\begin{figure}[h]
  \begin{center}
      \includegraphics[width=7in]{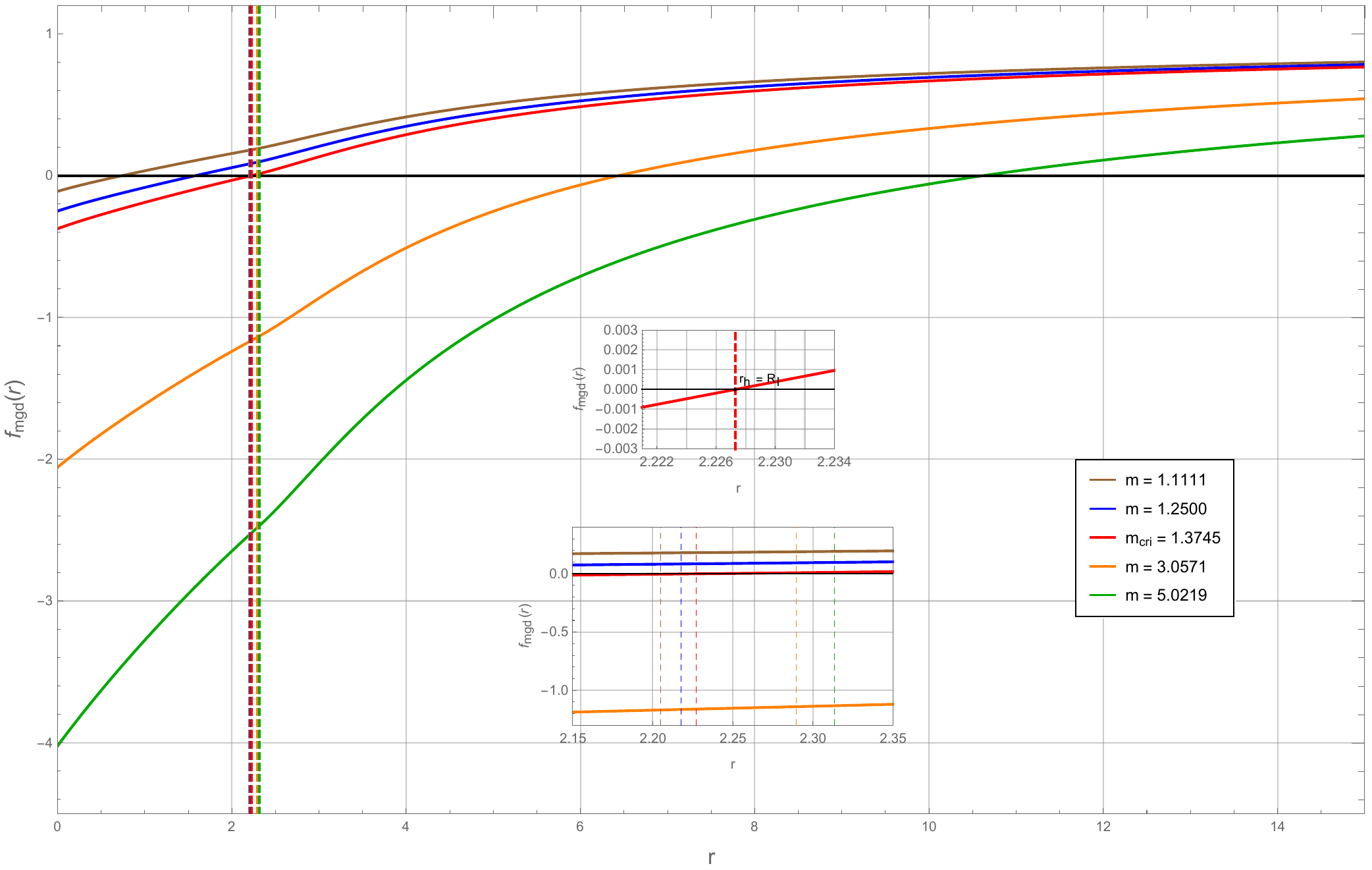}
            \caption{Metric function $f(r)$ versus the radial coordinate $r$, for $G_0=1$, $g_*=1$, $A=1$, $\gamma=9/2$, $\alpha=1$, $\rho_s=0.02$, $r_s=1$, and $K_I=0$, considering different values of the mass parameter $m=1.1111, 1.25, m_{\rm cri}=1.3745, 3.0571$, and $5.0219$. The values of $R_I$ are indicated by the vertical dashed lines.}
\label{FigfMGD}
  \end{center}
\end{figure}

Figure~\ref{FigfMGD} provides a complementary view through the complete metric function $f_{\rm mgd}(r)$ for representative values of the mass parameter. The vertical dashed lines indicate the corresponding matching radii $R_I$. The upper inset explicitly shows the critical configuration, for which the zero of $f_{\rm mgd}(r)$ coincides with the matching surface, while the lower inset illustrates the smooth behavior of the metric across the different matching radii. In all the representative configurations shown, the metric function exhibits a single zero. Therefore, within the numerical range investigated here, the inclusion of the effective NFW source does not generate an additional inner horizon and preserves the single-horizon structure sought in our construction.

We note that the running gravitational coupling converts the divergent NFW density into an effective source with a finite central density. In the ultraviolet regime, where $G_I(r)\sim r$, the $r^{-1}$ cusp of the NFW profile is softened and the effective density approaches a finite constant at the origin. In contrast, in the infrared regime, where $G_{II}(r)\to G_0$, the effective source asymptotically recovers the usual NFW behavior. The model therefore combines two distinct effects of the same running coupling: on the one hand, it governs the short-distance quantum-gravitational seed geometry; on the other hand, it gravitationally dresses the additional NFW matter distribution, softening its central cusp. The resulting additional matter sector is separately conserved and modifies the spacetime through the MGD mass function, without requiring a modification of the RG scale-setting prescription or of the matching radius associated with the running gravitational coupling.

\begin{table}[ht]
\centering
\caption{Representative numerical configurations obtained from the matching conditions for $G_0=1$, $g_*=1$, $A=1$, $\gamma=9/2$, $R_c=3.67049$, $\alpha=1$, $\rho_s=0.02$, $r_s=1$, and $K_I=0$.}
\begin{tabular}{cccccccc}
\hline\hline
$\xi$ &
$m$ &
$\xi_{II}$ &
$R_I$ &
$R_I/R_c$ &
$K_{II}$ &
$r_h$ &
$r_h/R_I$
\\
\hline
0.9900 & 1.0101 & 0.7001 & 2.1943 & 0.5978 & 0.01171 & 0.0621 & 0.0283 \\
0.9000 & 1.1111 & 0.6777 & 2.2050 & 0.6007 & 0.01189 & 0.7409 & 0.3360 \\
0.8000 & 1.2500 & 0.6501 & 2.2177 & 0.6042 & 0.01209 & 1.5834 & 0.7140 \\
0.7275 & 1.3745 & 0.6281 & 2.2273 & 0.6068 & 0.01225 & 2.2273 & 1.0000 \\
0.7170 & 1.3947 & 0.6247 & 2.2287 & 0.6072 & 0.01228 & 2.3189 & 1.0405 \\
0.6900 & 1.4494 & 0.6159 & 2.2325 & 0.6082 & 0.01234 & 2.5303 & 1.1334 \\
0.4718 & 2.1194 & 0.5313 & 2.2651 & 0.6171 & 0.01288 & 4.3254 & 1.9096 \\
0.3271 & 3.0571 & 0.4560 & 2.2896 & 0.6238 & 0.01328 & 6.4393 & 2.8124 \\
0.2479 & 4.0334 & 0.4041 & 2.3041 & 0.6277 & 0.01352 & 8.5398 & 3.7064 \\
0.1991 & 5.0219 & 0.3662 & 2.3136 & 0.6303 & 0.01368 & 10.6254 & 4.5926 \\
\hline\hline
\end{tabular}
\label{tab:matchingNFW}
\end{table}

\section{Discussion and Conclusion}

In this work, we have explored an alternative RG-improved BH geometry inspired by AS gravity, in which the UV structure is not determined by the requirement of complete regularity at the origin, a feature commonly associated with the emergence of an inner horizon and a potentially unstable dS core. Instead, we characterize the short-distance/high-energy regime through the radially weighted effective quantum-gravitational source entering the volume integrand of the quantum-corrected Ricci scalar. Specifically, the quantity $\sim r^2\,T_{\rm qc}(r)$ is used to measure the strength of the quantum-gravitational sector relevant for the RG scale identification, according to Eq.~\eqref{IdentificacionMia}. This prescription is motivated by the volume integrability of the Ricci scalar and leads to a qualitatively different UV structure from the one commonly associated with regular dS cores. Rather than removing the central divergence completely, the resulting geometry preserves it in an integrable form and, within the domain analyzed, exhibits a single event horizon without the additional inner horizon associated with a potentially unstable dS core.

A consequence of our identification is the emergence of a nontrivial radial dependence of the gravitational coupling in the UV region. In particular, we obtained a Bessel-type solution satisfying $G_I(r)\sim r$ near the origin. This behavior differs qualitatively from the scaling $G(r)\sim r^3$ that frequently appears in RG-improved BHs with a regular dS core. The UV solution also possesses a restricted physical domain determined by the conditions $G_I(r)<G_0$, $k^2(r)>0$, and the monotonicity of the coupling. Owing to these restrictions, together with the behavior of the Bessel functions beyond the allowed radial domain, we do not extrapolate this branch outside its regime of validity. Instead, we connect it through the Israel--Darmois junction conditions to an exterior RG-improved solution that recovers $G_0$ in the IR. As a representative test case, we choose the running coupling in Region II to take the Bonanno--Reuter form given in Eq.~\eqref{ExpresionG2Bonano} \cite{Bonanno:2000ep,Falls:2010he}. In this way, the construction combines an inner regime determined by the integrability condition with an exterior region exhibiting the expected gravitational behavior at large distances in the IR. The numerical analysis shows that, within the parameter space investigated, the resulting geometry possesses a single event horizon and does not develop the additional inner horizon commonly associated with a dS core. A critical configuration also appears for which the event horizon coincides with the boundary of Region I, $r_h=R_I$, separating solutions whose horizon lies in the inner Region I from those whose horizon belongs to the exterior Region II. Configurations with $r_h<R_I$ are particularly interesting, since the event horizon itself lies within the domain governed by the high-energy RG identification, which could be consistent with a Planckian scenario.

As a second physical problem, we considered the incorporation of an NFW dark matter distribution into the RG-improved geometry. BH solutions surrounded by NFW galactic halos have attracted increasing attention in recent years; see, e.g., Ref.~\cite{Xu:2018wow}, the references therein, and subsequent applications. In this context, it is natural to ask how the gravitational response of an NFW distribution is modified when the source is coupled to a running Newton coupling rather than to the constant value $G_0$. In our construction, the matter sector entering the gravitational equations is radially weighted by $G(r)$. This becomes particularly relevant in the UV, since the original NFW density behaves as $\rho_{\rm NFW}\sim r^{-1}$, whereas the running coupling satisfies $G_I(r)\sim r$ near the origin. Consequently, the corresponding effective matter sector entering the gravitational equations remains finite as $r\to0$, giving rise to a cored effective behavior at short distances, which we interpret in terms of an effective energy density. Importantly, this behavior is not introduced through an ad hoc modification of the NFW profile, but instead emerges directly at the level of the field equations from the gravitational coupling of the original distribution to $G(r)$. In the IR, where $G_{II}(r)\to G_0$, the conventional NFW behavior is progressively recovered. This result therefore establishes a direct connection between the UV quantum-gravitational structure of the BH geometry and the central gravitational response of the DM distribution. It is worth emphasizing that determining whether this mechanism can provide a complete description of the cusp--core problem at galactic scales lies beyond the scope of the present work.

To incorporate this additional source, we retained the same radial domains I and II and the assumptions underlying the RG scale identification adopted for the seed geometry. This choice follows from our interpretation that the scale setting is determined by the effective quantum-gravitational sector responsible for the running of $G(r)$, whereas the NFW distribution is introduced as an additional matter source that is gravitationally coupled to this already established running coupling. The incorporation of the NFW sector was carried out using the gravitational decoupling algorithm \cite{Ovalle:2017fgl}, which provides a controlled separation between the quantum-gravitational seed sector and the DM contribution. Both sectors remain separately conserved while contributing to the same geometry, consistently with the assumption that there is no direct exchange of energy and momentum between them and that their interaction is mediated gravitationally, in line with the usual assumption that DM does not directly interact with the ordinary matter sector through non-gravitational interactions. We also verified the Israel--Darmois junction conditions for the deformed geometry, obtaining that the additional matter contribution can be consistently matched across the same interface separating Regions I and II. The DM contribution also produces measurable geometric effects. For the configurations considered, it displaces the event horizon toward larger radial values and modifies the critical configuration separating solutions whose horizons lie in the inner and outer regions. Nevertheless, within the numerical domain investigated, this deformation does not generate an additional inner horizon: the metric function continues to exhibit a single zero. Therefore, the single-horizon structure that motivated the original construction is preserved even after the inclusion of the additional matter sector.

\appendix

\section{Brief Review of Asymptotically Safe Gravity}
\label{Apendice}

We shall understand $q$ as the momentum associated with a particular fluctuation of the quantum gravitational field, which can roughly be interpreted as a fluctuation of the metric, while $k$ denotes the energy scale of the renormalization group (RG). At a given scale $k$, the effective gravitational action is written as $\Gamma_k \sim G(k)^{-1} \int d^4x\,\sqrt{g}\left[-R(g)+\cdots\right]$, where $G(k)$ is the scale-dependent Newton coupling. In the ultraviolet limit, $k\to\infty$, $\Gamma_k$ approaches the microscopic or bare action, whereas for $k\to0$ it yields the full effective action. The flow between these limits is governed by the Wetterich equation \cite{Wetterich:1992yh},

\begin{equation}
\label{Flujo1}
k\partial_k\Gamma_k
=
\frac{1}{2}
\mathrm{STr}
\left[
\left(
\Gamma_k^{(2)}
+
\mathcal{R}_k
\right)^{-1}
k\partial_k\mathcal{R}_k
\right],
\end{equation}
where $\mathcal{R}_k$ is an infrared regulator, $\Gamma_k^{(2)}$ is the second functional derivative of the effective action, and $\mathrm{STr}$ denotes the supertrace over the relevant degrees of freedom. The regulator controls which momentum modes contribute to the flow. In particular, for $q\ll k$, $\mathcal{R}_k(q^2)\sim k^2$, strongly suppressing these modes. 
The connection between the Wetterich equation \eqref{Flujo1} and the dimensionless gravitational coupling $g(k)$ is obtained by projecting the flow of $\Gamma_k$ onto the term proportional to the Ricci scalar $R$. After the corresponding functional and algebraic treatment \cite{Platania:2023srt}, the running of the dimensionless Newton coupling can be studied by introducing
\begin{equation}
\label{Gk:001}
g(k)=k^2G(k).
\end{equation}
 
The evolution of $g(k)$ with the RG scale is determined by its beta function,

\begin{equation}
\label{beta:002}
\beta_g(g)
\equiv
k\frac{dg(k)}{dk}.
\end{equation}

A central assumption of asymptotically safe gravity is the existence of a finite and non-trivial ultraviolet fixed point $g_*$ approached by the RG flow as $k\to\infty$. At this fixed point, the beta function defined in Eq. \eqref{beta:002} satisfies $\beta_g(g_*)=0$, and the dimensionless coupling approaches a finite value, $\lim_{k\to\infty}g(k)=g_*$. The ultraviolet fixed-point behavior then implies

\begin{equation}
\label{GUV}
G(k)
\sim
\frac{g_*}{k^2}
\,\Rightarrow\,
G(k)\to0,
\,\,\,
k\to\infty.
\end{equation}

While the dimensionless gravitational coupling remains finite in the ultraviolet, the dimensionful Newton coupling vanishes asymptotically according to Eq. \eqref{GUV}. In the opposite limit, one expects to approximately recover the classical Newton coupling in the infrared regime, $\displaystyle \lim_{k\to0}G(k)=G_0$. Therefore, there are two limiting behaviors of the running Newton coupling: the classical gravitational coupling is recovered in the IR, whereas $G(k)$ decreases as $k^{-2}$ near the ultraviolet fixed point. The equation describing the flow of the dimensionless gravitational coupling is particularly convenient, since it allows us to obtain a simple explicit expression for $G(k)$ once a reference condition is specified. For example, taking the infrared limit $k_0\to0$ as a reference, where $G(k_0)=G_0$, one obtains
\begin{equation}
\label{Gkflow}
G(k)
=
\frac{G_0}
{1+g_*^{-1}G_0 k^2}.
\end{equation}

This equation was derived in Ref. \cite{Bonanno:2000ep}] from an appropriate algebraic treatment of Eq. \eqref{Flujo1}. At low-energy scales, $k\ll1$, Eq.~\eqref{Gkflow} approximately gives $G(k)\simeq G_0$, thus recovering the classical behavior in the infrared regime. On the other hand, at high-energy scales, $k\gg1$ Eq.~\eqref{Gkflow}, so that $G(k)\simeq \frac{g_*}{k^2}$, reproducing the characteristic ultraviolet behavior near the fixed point.

\section{Analytical Solution for the Mass Function in Region II}
\label{AppendixMassII}

In this appendix, we briefly present the analytical integration of Eq.~\eqref{MasaAlphaIIIntegral}. The denominator contains the cubic polynomial
\begin{equation}
\mathcal{P}_{II}(r)
=
r^3+
\frac{\xi_{II}^{\,2}}{g_*}G_0
(r+\gamma mG_0).
\end{equation}

For the positive parameters considered in our model, this polynomial possesses a unique negative real root $q$, determined by $\mathcal{P}_{II}(q)=0$. For $q\neq-r_s$, the analytical solution is given by
\begin{equation}
\label{MasaAlphaIIAnalytical}
\begin{aligned}
m_{\alpha,II}(r)
={}&4\pi G_0\rho_s r_s^3
\Bigg[
C_1\ln(r+r_s)
-\frac{C_2}{r+r_s}+B\ln(r-q)+\frac{c}{2}
\ln\left(
r^2+qr+q^2+
\frac{\xi_{II}^{\,2}G_0}{g_*}
\right)\\
&+
\frac{d-cq/2}
{\sqrt{
\displaystyle\frac{\xi_{II}^{\,2}G_0}{g_*}
+\frac{3q^2}{4}
}} \times
\arctan\left(
\frac{r+q/2}
{\sqrt{
\displaystyle\frac{\xi_{II}^{\,2}G_0}{g_*}
+\frac{3q^2}{4}
}}
\right)
\Bigg]+K_{II}.
\end{aligned}
\end{equation}
where the coefficients are given by
\begin{equation}
C_1=
\frac{
r_s^3\left[
r_s^3+
\displaystyle\frac{\xi_{II}^{\,2}G_0}{g_*}
(3r_s-4\gamma mG_0)
\right]
}{
\left[
\displaystyle\frac{\xi_{II}^{\,2}G_0}{g_*}
(\gamma mG_0-r_s)-r_s^3
\right]^2
},
\end{equation}
\begin{equation}
C_2=
\frac{r_s^4}{
\displaystyle\frac{\xi_{II}^{\,2}G_0}{g_*}
(\gamma mG_0-r_s)-r_s^3
},
\end{equation}
\begin{equation}
B=
\frac{q^4}{
(q+r_s)^2
\left(
3q^2+\displaystyle\frac{\xi_{II}^{\,2}G_0}{g_*}
\right)
},
\end{equation}
and
\begin{equation}
c=1-C_1-B,
\qquad
d=-2r_s+r_sC_1-C_2-qB+qc.
\end{equation}

In the particular case $q=-r_s$, corresponding to a repeated factor in the denominator, the partial fraction decomposition must be modified to account for a pole of order three. Nevertheless, the integral remains analytically solvable in terms of logarithmic, rational, and inverse trigonometric functions. Thus, the mass function $m_{\alpha,II}(r)$ admits an exact analytical representation for all positive values of the model parameters, although its explicit form depends on the multiplicity of the polynomial roots.

\bibliography{mybib.bib}

\end{document}